\documentclass{elsart}

\usepackage{graphicx}

\usepackage{amssymb}
\DeclareMathAlphabet{\mathsfsl}{OT1}{cmss}{m}{sl}

\begin{document}

\newcommand{\tensor}[1]{\mathsfsl{#1}}

\newcommand{\br}{{\bf r}}
\newcommand{\f}{{\bf f}}
\newcommand{\that}{\hat {\bf t}}
\newcommand{\nhat}{\hat {\bf n}}
\newcommand{\bhat}{\hat {\bf b}}
\newcommand{\ehat}{\hat{\bf e}}
\newcommand{\xhat}{\hat{\bf x}}
\newcommand{\yhat}{\hat{\bf y}}
\newcommand{\zhat}{\hat{\bf z}}
\newcommand{\rhat}{\hat{\bf r}}
\newcommand{\bF}{{\bf F}}
\newcommand{\phihat}{\mbox{\boldmath$\hat\varphi$}}
\newcommand{\thetahat}{\mbox{\boldmath$\hat\theta$}}
\newcommand{\rhohat}{\mbox{\boldmath$\hat\rho$}}
\newcommand{\bOm}{\mbox{\boldmath$\Omega$}}
\newcommand{\bom}{\mbox{\boldmath$\omega$}}
\newcommand{\bep}{\mbox{\boldmath$\epsilon$}}
\newcommand{\bK}{{\bf K}}
\newcommand{\bN}{{\bf N}}
\newcommand{\bM}{{\bf M}}

\begin{frontmatter}



\title{Dynamic Supercoiling Bifurcations of Growing Elastic Filaments}


\author[cww]{Charles W. Wolgemuth}
\ead{cwolgemuth@uchc.edu}
\address[cww]{Department of Cell Biology, University of Connecticut Health
Center,
Farmington, CT 06030}

\author[reg]{Raymond E. Goldstein}
\ead{gold@physics.arizona.edu}
\address[reg]{Department of Physics and Program in Applied
Mathematics, University of Arizona, Tucson, AZ  85721}

\author[trp]{Thomas R. Powers}
\ead{Thomas\_Powers@brown.edu}
\address[trp]{Division of
Engineering, Brown University, Providence, RI 02912}

\begin{abstract}
Certain bacteria form filamentous colonies when the cells fail to
separate after dividing. In {\it Bacillus subtilis}, {\it Bacillus
thermus}, and {\it cyanobacteria}, the filaments can wrap into
complex supercoiled structures as the cells grow. The structures
may be solenoids or plectonemes, with or without branches in the
latter case. Any microscopic theory of these morphological
instabilities must address the nature of pattern selection in the
presence of {\it growth}, for growth renders the problem
nonautonomous and the bifurcations dynamic. To gain insight into
these phenomena, we formulate a general theory for growing elastic
filaments with bending and twisting resistance in a viscous
medium, and study an illustrative model problem: a growing
filament with preferred twist, closed into a loop. Growth depletes
the twist, inducing a twist strain. The closure of the loop
prevents the filament from unwinding back to the preferred twist;
instead, twist relaxation is accomplished by the formation of
supercoils. Growth also produces viscous stresses on the filament
which even in the absence of twist produce buckling instabilities.
Our linear stability analysis and numerical studies reveal two
dynamic regimes. For small intrinsic twist the instability is akin
to Euler buckling, leading to solenoidal structures, while for
large twist it is like the classic writhing of a twisted filament,
producing plectonemic windings. This model may apply to situations
in which supercoils form only, or more readily, when axial
rotation of filaments is blocked. Applications to specific
biological systems are proposed.

\end{abstract}

\end{frontmatter}

\section{Introduction}
\label{intro}

Many structures in the biological world grow so slowly that they
adopt a shape that can be considered as a minimizer of some
configurational energy associated solely with the {\it internal}
structure. The logarithmic spiral of the nautilus shell is an
example. It enlarges through a process of differential growth
whereby its shape represents the accumulated history of identical
events, save for scale changes~\cite{thompson1942}.  Thus, the
microscopic rules of growth are essentially unchanging as the
three-dimensional form develops, and the properties of the
external environment do not fundamentally determine the form.

Environmental effects on the development of biological forms are
well known.  Consider the formation of tendril perversions in
climbing vines~\cite{Darwin,goriely_tabor1998}.  A perversion is a
junction between regions of opposite helix handedness that forms
as an initially straight tendril first attaches to a support
structure and then, through the activation of tension-sensitive
receptors, undergoes a helical instability. The constraint of
fixed ends enforces the formation of a structure with zero net
twist, consisting of concatenated regions of opposite chirality
joined by a transition region---the perversion. Thus, the
interaction with the external environment fundamentally alters the
pattern formation through the tension induced by a point contact.

Even more complex phenomena occur for those structures whose very
process of growth induces an {\it external}
force, as with motion through an environment
that offers viscous resistance.  These new forces can dramatically
alter the ultimate configuration, and the present paper is a case study
in such phenomena.  We focus on a system in which the formation of
patterns occurs through a finite-wavelength instability in which
the process of growth introduces an intrinsic time dependence to the
control parameters.  The resulting bifurcation problem is nonautonomous
and can exhibit a rich phenomenology as the intrinsic time scale of
growth competes with those of the various modes of instability.
This competition places the problem among the class of so-called
``dynamic bifurcations," of which many examples are of
continuing interest.  These include instabilities in directional
solidification in which the initial
acceleration of the interface from rest provides the nonautonomous
character \cite{warren_langer}, fingering instabilities of magnetic fluids
under the influence of time-dependent magnetic fields
\cite{ferro_science,ferro_pre}, and separatrix crossing in Hamiltonian
systems viewed as models for stellar fission \cite{pesci_lebovitz}.

Motivation for this focus comes from the phenomenon of
supercoiling exhibited by filamentous colonies of {\it Bacillus
subtilis} and other bacteria. {\it B. subtilis} cells are
rod-shaped bacteria, typically four microns in length and slightly
over one-half micron in diameter. Wild-type rod-shaped bacteria
grow by extending along the cylindrical axis of symmetry, and then
dividing and separating in the middle~\cite{mendelson82,koch2000}.
Under certain conditions, the cells of some mutant forms fail to
separate upon replication, leading to a long chain of cells. Other
species have also been observed to form chains, including {\it
Escherichia coli}~\cite{e_coli}, {\it Cyanobacteria}~\cite{wolk},
{\it Myxococcus xanthus}~\cite{sun}, and {\it Mycobacterium
tuberculosis}~\cite{tuberculosis}. Under certain growth
conditions, strains of {\it B. subtilis}, {\it Bacillus
stearothermophilus}~\cite{stearotherm}, {\it
Thermus}~\cite{thermus}, and {\it Mastigocladus
laminosus}~\cite{laminosus} form complex braided structures.
Helical (or, in the jargon of DNA biophysics, ``solenoidal")
morphologies have also been observed~\cite{tilbey1977}. Of these
examples, the supercoiled structures of {\it B. subtilis} have
been studied most extensively~\cite{mendelson76,mendelson95}.
Recent experimental work \cite{Jones,BenYehuda} has indicated that
the cell wall of {\it B. subtilis} contains helical protein
structures.  These may supply the molecular imprinting responsible
for this morphological development, in a manner analogous to the
way microtubules control macroscopic handedness in certain plants
\cite{arabidopsis}, but as of yet no successful microscopic theory
for the formation of these supercoiled structures exists.

\begin{figure}
\includegraphics[height=3in]{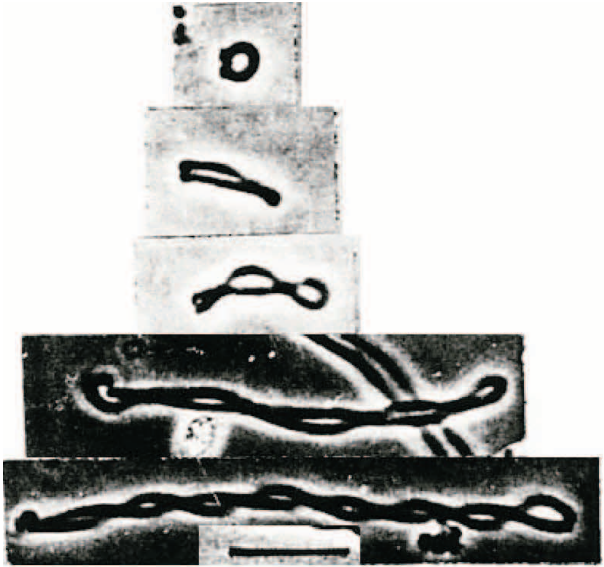}
\caption{Phase contrast micrographs of growing {\it B. subtilis}
spores with attachment of both cell poles to the spore coat.  The
series consists of different filaments at various stages of
growth. Scale bar = $10$ $\mu$m. Figure courtesy of N. Mendelson;
see~\cite{mendelson76}. } \label{expt_loop}
\end{figure}

Throughout much of the development of complex structures in {\it
B. subtilis}, the length $L$ of the elongating chain, a single
cell thick, grows exponentially in time, $L\propto\exp(\sigma t)$,
with the growth rate $\sigma\approx2\times10^{-4}$ s$^{-1}$.
Moreover, material cross-sections of the cells rotate relative to
each other \cite{Neil_recent}, so that the angle describing the
relative orientation of any two material cross-sections is
proportional to their exponentially increasing separation. In one
of the earliest observations of supercoiling in {\it B. subtilis},
the ends of the fiber adhered to a spore coat, prohibiting axial
rotation of the ends~\cite{mendelson76}. Mendelson supposed that
this blocked rotation leads to torsional stress, eventually
causing supercoiling~\cite{mendelson76} (See Fig.
\ref{expt_loop}). Since this discovery, it has been shown that
adhesion is not required for supercoiling \cite{mendelson95}, and
that many different factors such as temperature, pH, and the
concentration of ions such as magnesium and ammonium affect the
morphology of the coils~\cite{temperature,ion_conc}. The evolution
of the coils after the formation of the first braid is remarkable.
The first plectonemic braid, essentially a filament that is two
cells thick, continues to grow and eventually reaches a critical
length of order 100 $\mu$m, after which it supercoils to form
another braided structure which is four cells thick.  This process
can repeat many times, leading to a hierarchy of braids and
eventually a macroscopic object.

In this paper, we study in detail Mendelson's blocked
rotation mechanism of supercoiling in a growing closed loop, and focus on
the formation of the initial braid.  While not yet explaining the
microscopic physical origin of the coiling instability, we elucidate the
rich dynamics that occurs when growth competes with blocked rotation, and
thereby help constrain more detailed theories.

With this focus we exploit several simplifying assumptions in our
analysis. First, the chain of cells is treated as an elastic
filament with uniform properties along its length. There is
evidence that this assumption holds until times comparable to a
few doubling times ($\sim\sigma^{-1}$), but may be violated later.
For example, when the chirality of nutrient molecules in the
growth medium is reversed, the supercoils unwind and even begin to
wrap up in the opposite handedness, but the hairpin bends from the
original braid remain~\cite{neil}. Thus, some of the deformation
of the growing filament becomes permanent.  This phenomenon is
reminiscent of the morphological development of plant tendrils, in
which young and flexible tendrils age with time, becoming woody
and locked in a fixed shape ~\cite{larson2000}. On the time scales
that will concern us, $t<\lesssim\sigma^{-1}$, single fibers have
been shown to behave like elastic rods, with a bending modulus
$A=10^{-12}$ erg-cm~\cite{walk}, as discussed further in
section~\ref{constit}.

A second major simplification we introduce is to treat the growth
rate as constant in time, independent of stress and filament
geometry. Again, the permanent hairpin bends in the chirality
reversal experiment show that this assumption cannot hold
everywhere along the filament for all times after the first braid
forms.

Together with the observation that viscous effects dominate
inertial effects in the low Reynolds number environment of the
growing fibers, these assumptions lead to the model studied in
this paper: an elastic ring with intrinsic twist suspended in a
viscous fluid and lengthening at an exponential rate. Our work is
complementary to that of other investigators. For example, Shelley
and Ueda studied the Euler-like buckling of a growing liquid
crystal filament, using a local drag model for the linear
stability of a growing loop~\cite{shelleyUeda}, and incorporating
nonlocal Stokesian hydrodynamics to study the pattern
formation~\cite{shelley}. Drasdo studied similar patterns in the
context of the growth of single-cell-layer tissue
sheets~\cite{drasdo}.  Klapper has studied inertial writhing
instabilities of {\it open} rods subject to exponential growth, as
well as the relaxation to equilibrium of twisted rings in the
absence of growth~\cite{klapper}.  Goriely and
Tabor~\cite{goriely2000} introduced the idea of {\it twist
depletion} as a possible mechanism driving buckling in {\it B.
subtilis}.

Our analysis begins with a generalization of the kinematics and
dynamics of slender filaments to account for growth.
Section~\ref{grow_elas_loop} treats the growing elastic loop,
beginning with a qualitative discussion of the instability. The
linear stability analysis is greatly simplified by the use of the
natural frame, so in this section we include a self-contained
summary of the properties of the natural frame. We present a
quasi-analytic treatment of the linear stability of the loop, and
then present numerical simulations of the full nonlinear problem.
Section~\ref{conclude} is the conclusion.

\section{Kinematics and Dynamics of Growing Rods}

\subsection{Centerline Kinematics}

In this section we extend the standard kinematics and dynamics of
elastic rods to allow for growth.  Let $s$ denote the arclength
measured at time $t$ from one end of an open rod, or from a fixed
material point for a rod closed to form a loop. Then
$\mathbf{r}(s,t)$ is the position in space of the centerline of
the rod with arclength coordinate $s$ at time $t$.  Since material
points on the rod centerline are convected along the rod by
growth, fixed values of $s$ do not correspond to fixed material
points. We choose to label the material points of the centerline
at all times by the arclength parameterization $s_0$ at a fixed
time $t=0$. We will study exponential growth, for which
$s=\exp(\sigma t)s_0$. Note that the partial derivatives with
respect to $s$ and $t$ commute because $s$ and $t$ are independent
variables:
\begin{equation}
\left. {\partial\over\partial t}\right|_s
\left. {\partial\over\partial s}\right|_t=
\left. {\partial\over\partial s}\right|_t
\left. {\partial\over\partial t}\right|_s;
\label{commute}
\end{equation}
however, $\partial/\partial t|_{s_0}$ and $\partial/\partial s|_t$
do not commute since
\begin{equation}
\left. {\partial\over\partial t}\right|_{s_0}= \left.
{\partial\over\partial t}\right|_s+ \left. {\partial
s\over\partial t}\right|_{s_0} \left. {\partial\over\partial
s}\right|_t, \label{donotcommute}
\end{equation}
by the chain rule.

The
velocity of a material point is the time derivative of position at
fixed $s_0$,
\begin{equation}
\mathbf{v}(s_0,t)=\left. {\partial\mathbf{r}\over\partial t}
\right|_{s_0}.
\label{v-eqn}
\end{equation}
It is convenient to develop the equations of motion in terms of
$s$ rather than $s_0$, so that
\begin{equation}
\mathbf{v}(s,t)=\left. {\partial\mathbf{r}\over\partial t}
\right|_s+\left.
{\partial s\over \partial t}\right|_{s_0}\left.
{\partial\mathbf{r}\over\partial s}\right|_t.
\label{convect1}
\end{equation}
To avoid confusion, we will explicitly denote which variables are
fixed when finding the partial derivatives with respect to $t$. However,
since partial derivatives with respect to $s$ will always be taken
at fixed $t$, we will write $\partial/\partial s|_t=\partial/\partial
s$.

The first term of Eq.~(\ref{convect1}) corresponds to the velocity
of the centerline in the absence of growth, and the second term
arises from growth-induced convection.
Henceforth we will specialize to exponential growth,
for which $\partial s/\partial t|_{s_0}=\sigma s$ and
\begin{equation}
\mathbf{v}=\left. {\partial\mathbf{r}\over\partial t}\right|_s
+s\sigma{\partial\mathbf{r}\over\partial s}=\left.
{\partial\mathbf{r}\over\partial t}\right|_s+s\sigma
 \mathbf{\hat e}_3,
\label{convect2}
\end{equation}
where $\mathbf{\hat e}_3=\partial\mathbf{r}/\partial s$ is the
unit tangent vector of the centerline. Although the velocity of a
material point may seem from Eq.~(\ref{convect2}) to depend on the
arbitrary choice of origin for $s$, Eq.~(\ref{v-eqn}) shows that
the velocity is manifestly independent of this choice. Note that
Eq.~(\ref{convect2}) implies that $\mathbf{\hat
e}_3\cdot\partial\mathbf{v}/\partial s=\sigma$.

\begin{figure}
\includegraphics[height=2in]{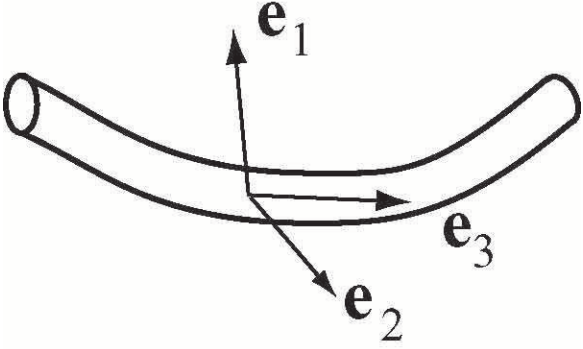}
\caption{The material frame.  The unit vector $\mathbf{\hat e}_3$
is
tangent to the rod centerline. The other members of the orthonormal
frame, $\mathbf{\hat e}_1$ and
$\mathbf{\hat e}_2$, point to material points on the rod's surface.}
\label{mat_frame}
\end{figure}

\subsection{Choice of Frame and Growth Model}
\label{frame_growth_model}

In Kirchhoff rod theory~\cite{kirchhoff}, the configuration of a
rod is completely specified by the orientation of the material
orthonormal frame $\{\mathbf{\hat e}_1,\mathbf{\hat
e}_2,\mathbf{\hat e}_3\}$. The vectors $\mathbf{\hat e}_1$ and
$\mathbf{\hat e}_2$ point to material points on the rod surface
(Fig.~\ref{mat_frame}). As the rod bends and twists, the positions
of these material points change, causing the material frames to
rotate~\cite{landau}:
\begin{eqnarray}
{\partial \mathbf{e}_i\over\partial s}&=&\mathbf{\Omega}
\times\mathbf{e}_i\label{space_rot}\\
\left. {\partial \mathbf{e}_i\over\partial t}
\right|_{s_0}&=&\bom\times\mathbf{e}_i.
\label{time_rot}
\end{eqnarray}
The vector $\mathbf{\Omega}$ describes the bending and twisting
strain at a given instant, and $\bom$ is the angular velocity of a
material frame at a given material point $s_0$.  In general,
$\mathbf{\Omega}$ and $\bom$ depend on the choice of material
frame. Once the choice is made for a given configuration, say the
stress-free state, then the choice is specified for all
configurations. In the classical rod theory without growth, it is
natural to align $\mathbf{\hat e}_1$ and $\mathbf{\hat e}_2$ with
the principal axes of the cross-section.  If the cross-section is
circular, as we henceforth assume, and if the rod is straight in
the absence of stress, then there are many equivalent natural
choices. For example, if the rod aligns along the $z$-axis when it
is stress-free, then $\{\mathbf{\hat e}_1,\mathbf{\hat
e}_2\}=\{\mathbf{\hat x}, \mathbf{\hat y}\}$ is natural. Any
uniform rotation of this frame about $\mathbf{\hat z}$ is equally
convenient; all these choices lead to $\mathbf{\Omega}=0$ in the
absence of stress.

If the rod is curved in the stress-free state, then the direction
of curvature breaks the rotational symmetry of the circular
cross-section and provides a natural choice for the directions of
$\mathbf{\hat e}_1$ and $\mathbf{\hat e}_2$.  {\it E.g.}, if the
rod has a helical stress-free state, then we may take
$\mathbf{\hat e}_1=\mathbf{\hat n}$ and $\mathbf{\hat e}_2
=\mathbf{\hat b}$, where $\mathbf{\hat n}$ and $\mathbf{\hat b}$
are the unit normal and binormal of the Frenet-Serret
frame~\cite{Frenet,serret,kamienrmp}:
\begin{eqnarray}
{\partial\ehat_3\over\partial s}&=&\kappa\nhat\label{F-S1}\\
{\partial\nhat\over\partial s}&=&-\kappa\ehat_3+\tau\bhat\label{F-S2}\\
{\partial\bhat\over\partial s}&=&-\tau\nhat,\label{F-S3}
\end{eqnarray}
where $\kappa$ is the curvature and $\tau$ the torsion.
  Thus,
$\mathbf{\Omega}=\kappa_0\mathbf{\hat e}_2+\tau_0\mathbf{\hat
e}_3$ for this frame in the absence of stress, where $\kappa_0$ is
the curvature and $\tau_0$ is the torsion of the helix.

\begin{figure}
\includegraphics[height=3in]{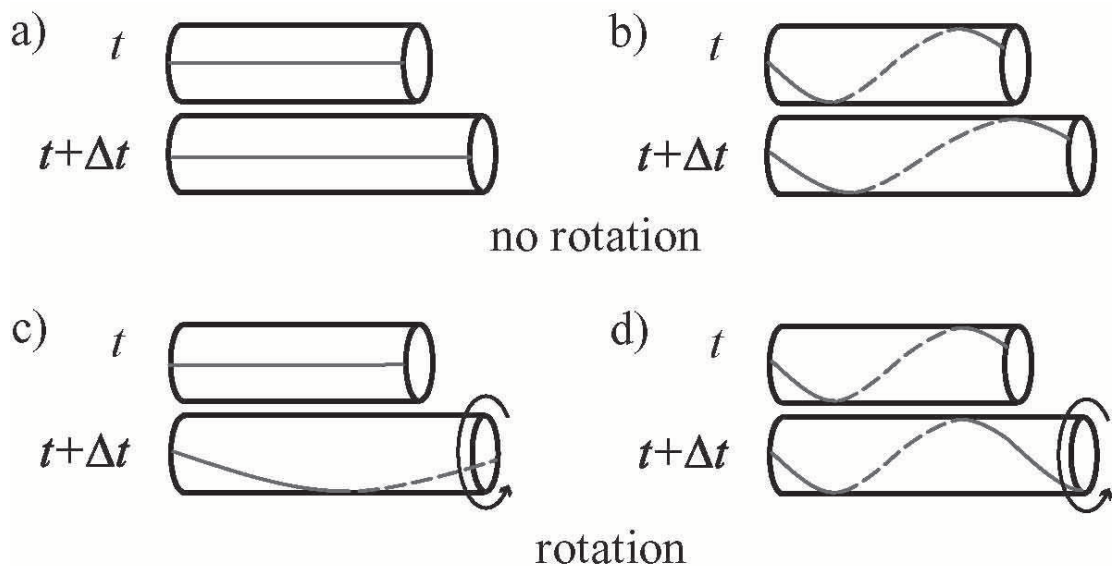} \caption{Two different modes of
growth. In (a) and (b), the rod lengthens without rotation. In (c)
and (d), cross-sections at fixed material points rotate with an
ever-increasing angular velocity.} \label{growth_fig}
\end{figure}

When $\kappa_0=0$, the helix of the last example degenerates to a
straight rod with spontaneous twist $\tau_0$. We will show below
that if $\kappa_0=0$ and the cross-section is circular, then
$\tau_0$ can be eliminated from the equations of motion for an
inextensible (non-growing) rod, and therefore does not affect the
rod shape and dynamics. However, $\tau_0$ has physical meaning for
a {\it growing} rod, even if $\kappa_0=0$ and the cross-section is
circular. Figure~\ref{growth_fig} illustrates two kinematic
possibilities for growth. Each sub-figure shows the stress-free
configuration of a growing rod at two different times. In all
cases, the left end of the rod has a fixed position and
orientation.  Two growth schemes are shown; in the growth scheme
of Fig.~\ref{growth_fig}a and~\ref{growth_fig}b, the material
frames are carried to greater values of $z$ by growth and have no
angular velocity. A line of material points parallel to the
$z$-axis at time $t$ remains parallel to the $z$-axis at time
$t+\Delta t$ (Fig.~\ref{growth_fig}a). However, the pitch of a
{\it helical} line of material points increases as the filament
grows (Fig.~\ref{growth_fig}b). Since the angular velocity is
zero, the number of helical turns is constant.

Figures~\ref{growth_fig}c and \ref{growth_fig}d illustrate a
growth scheme in which the cross-sections rotate with an angular
velocity that increases with arclength. In this case, a line of
material points parallel to the $z$-axis at time $t$ wraps around
the rod at time $t+\Delta t$ (Fig.~\ref{growth_fig}c). As the rod
grows, the material frames are carried to greater values of $z$
but also {\it rotate} relative to the fixed frame at $s=0$. In
this paper, we will study the growth model of
Fig.~\ref{growth_fig}c and d since it describes the relative
rotation of the cross-sections of the {\it B. subtilis}
fibers~\cite{fein1980,koch1990}. Define
$\theta(s_0)=\cos^{-1}(\mathbf{\hat e}_1(0)\cdot \mathbf{\hat
e}_1(s_0))$ as the angle between the orientation of the
(zero-stress) material frames at $s_0$ and $s=0$. We will suppose
that $\theta(s_0)$ increases linearly with $s_0$ and exponentially
in time with rate $\sigma$:
\begin{equation}
\theta(s_0)=\tau_0s_0\exp(\sigma t)=\tau_0 s.
\label{thetaeq}
\end{equation}
Since $\theta(s_0,t)$ and $s(s_0,t)$ increase in time with the
same exponential rate, a helical material line on the rod surface
with pitch $2\pi/\tau_0$ remains a helical line with the same
pitch as time passes (Fig.~\ref{growth_fig}d).  Thus, the natural
choice for the material frame in the stress-free state is
\begin{eqnarray}
\mathbf{\hat e}_1&=&\cos(s\tau_0)\mathbf{\hat
x}+\sin(s\tau_0)\mathbf{\hat y}\label{e1grow}\\
\mathbf{\hat e}_2&=&-\sin(s\tau_0)\mathbf{\hat
x}+\cos(s\tau_0)\mathbf{\hat y}\label{e2grow}\\
\mathbf{\hat e}_3&=&\mathbf{\hat z}.\label{e3grow}
\end{eqnarray}
Note that $\mathbf{\Omega}=\tau_0\mathbf{\hat e}_3$ and
$\bom=\sigma s\tau_0\mathbf{\hat e}_3$ for this frame; $\bom\neq0$
since the material frames must {\it rotate} to maintain zero
stress as the rod grows (in contrast to the growth model of Fig.
(a)). The parameter $\tau_0$ not only characterizes the
configuration of the natural material frame (as in the
inextensible helix example) but {\it also} the mode of growth.

\subsection{Compatibility relations}
\begin{figure}
\includegraphics[height=2.718in]{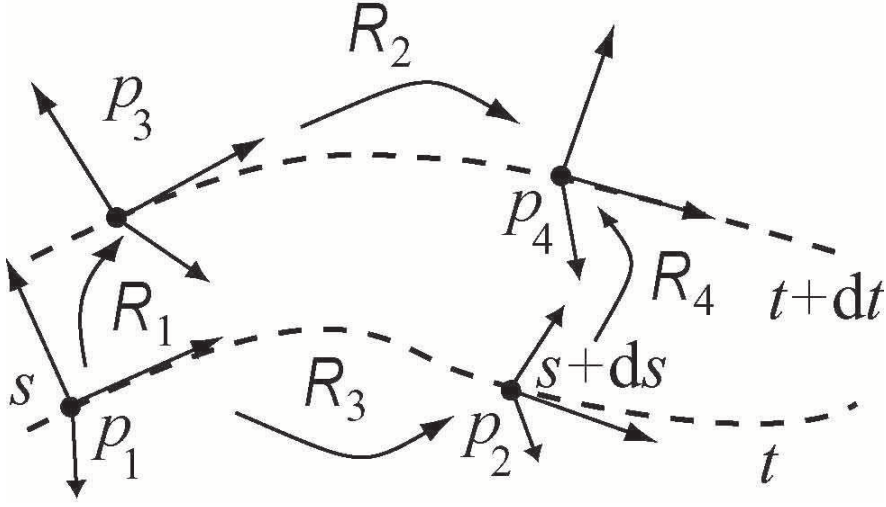}
\caption{Geometrical origin of compatibility relations.}
\label{frames_diagram}
\end{figure}

Geometry relates the strain vector $\mathbf{\Omega}$ and the
angular velocity $\bom$.  To see how, consider
Fig.~\ref{frames_diagram}. The lower curve represents the
centerline of the rod at time $t$, and the upper curve represents
the centerline of the rod at time $t+\mathrm{d}t$. The labelled
points on the upper curve have the same material coordinate as the
corresponding points on the lower curve; thus, the arclength
parameter for  $p_3$ is $s\exp(\sigma\mathrm{d}t)\approx
s+s\sigma\mathrm{d}t$, while the arclength parameter for $p_4$ is
$(s+\mathrm{d}s)\exp(\sigma\mathrm{d}t)\approx (s+\mathrm{d}s)
(1+\sigma\mathrm{d}t)$. Let $\tensor{R}_1$ be the rotation matrix
carrying the frame at $p_1$ to $p_3$, $\tensor{R}_2$ the rotation
matrix carrying the frame at $p_3$ to $p_4$, $\tensor{R}_3$ the
rotation matrix carrying the frame at $p_1$ to $p_2$, and
$\tensor{R}_4$ the rotation matrix carrying the frame at $p_2$ to
$p_4$.  Furthermore, let $\tensor{J}$ and $\tensor{K}$ denote the
infinitesimal rotation matrices associated with the rotation
vectors $\mathbf{\Omega}$ and $\bom$ respectively ({\it e.g.}
$\tensor{J}_{\alpha\beta}=
\epsilon_{\alpha\beta\gamma}\mathbf{\Omega}_\gamma$, where
$\epsilon_{\alpha\beta\gamma}$ is the alternating symbol). {}From
the definitions of the rotation matrices,
$\tensor{R}_2\tensor{R}_1 =\tensor{R}_4\tensor{R}_3$, or
\begin{eqnarray}
\Big[\tensor{I}+(1+\sigma\mathrm{d}t)\mathrm{d}s\tensor{J}
(s&+&s\sigma\mathrm{d}t,t+\mathrm{d}t)\Big]
\Big[\tensor{I}+\mathrm{d}t\tensor{K}(s,t)\Big]\nonumber\\
\approx\Big[\tensor{I}&+&\mathrm{d}t\tensor{K}(s+\mathrm{d}s,t)\Big]
\Big[\tensor{I}+\mathrm{d}s\tensor{J}(s,t)\Big],\label{rotations}
\end{eqnarray}
where $\tensor{I}$ is the identity matrix.
Expanding Eq.~(\ref{rotations}) to $\mathcal{O}
(\mathrm{d}s\mathrm{d}t)$, we find
\begin{eqnarray}
{\partial\tensor{K}\over\partial s}&=&\left. {\partial\tensor{J}
\over\partial t}\right|_s+\sigma s{\partial\tensor{J}\over\partial
s}
+\sigma\tensor{J}+[\tensor{J},\tensor{K}]\label{rotations2a}\\
{\partial\tensor{K}\over\partial s}&=&\left. {\partial\tensor{J}
\over\partial t}\right|_{s_0}
+\sigma\tensor{J}+[\tensor{J},\tensor{K}], \label{rotations2}
\end{eqnarray}
where $[,]$ is the commutator.  In terms of components in the material
frame,
\begin{eqnarray}
\left. {\partial\Omega_1\over\partial t}
\right|_{s_0}&=&{\partial\omega_1\over\partial s}-\sigma\Omega_1
+\Omega_2\omega_3-\Omega_3\omega_2\label{comp1}\\
\left. {\partial\Omega_2\over\partial t}
\right|_{s_0}&=&{\partial\omega_2\over\partial s}-\sigma\Omega_2
+\Omega_3\omega_1-\Omega_1\omega_3\label{comp2}\\
\left. {\partial\Omega_3\over\partial t}
\right|_{s_0}&=&{\partial\omega_3\over\partial s}-\sigma\Omega_3
+\Omega_1\omega_2-\Omega_2\omega_1.
\label{comp3}
\end{eqnarray}
These compatibility relations show how strain changes in time due
to non-uniform rotation rates (the first term in each of
Eqs.~(\ref{comp1})--(\ref{comp3})), growth (the second term in
each of Eqs.~(\ref{comp1})--(\ref{comp3})), and the geometric
coupling between twisting and bending (the last two terms in each
of Eqs.~(\ref{comp1})--(\ref{comp3})). The compatibility relation
Eq.~(\ref{comp3}) will be used below to determine the dynamics of
the twist strain $\Omega_3$.  Note that Eq.~(\ref{comp3}) can be
re-written in a form valid for arbitrary growth
laws~\cite{twirl,bistable_helices,klapper94}
\begin{equation}
\left. {\partial\Omega_3\over\partial
t}\right|_{s_0}={\partial\omega_3\over\partial s}
-\Omega_3{\partial\mathbf{r}\over\partial s}\cdot{\partial
\mathbf{v}\over\partial s}
+{\partial\mathbf{r}\over\partial
s}\times{\partial^2\mathbf{r}\over\partial s^2}\cdot
{\partial \mathbf{v}\over\partial s}.
\end{equation}

\subsection{Constitutive relations}
\label{constit}
The Kirchhoff constitutive relations for a rod with intrinsic
twist relate the moment on a cross-section of the
rod to the strain~\cite{love}:
\begin{equation}
\mathbf{M}=A\bigg({\partial\mathbf{r}\over\partial s}
\times{\partial^2\mathbf{r}\over\partial s^2} \bigg)
+C(\Omega_3-\tau_0)\mathbf{\hat e}_3. \label{khirch_const}
\end{equation}
Note that $\partial\mathbf{r}/\partial s
\times\partial^2\mathbf{r}/\partial s^2=\Omega_1\mathbf{\hat
e}_1+\Omega_2\mathbf{\hat e}_2$. %
Equation~(\ref{khirch_const}) implies that in
the stress-free state (defined by $\mathbf{M}=0$), $\mathbf{\hat
e}_1$ and $\mathbf{\hat e}_2$ rotate around $\mathbf{\hat e}_3$
with rate $\tau_0=\mathbf{\hat e}_2\cdot\partial \mathbf{\hat
e}_1/\partial s$, and $\mathbf{\hat e}_3$ is constant.

The force and moment balances for a growing rod are the same as
the balances for an inextensible rod:
\begin{eqnarray}
{\partial\mathbf{F}\over\partial
s}+\mathbf{f}_{\mathrm{ext}}&=&0\label{fbal}\\
{\partial\mathbf{M}\over\partial s}+\mathbf{\hat e}_3\times\mathbf{F}
+\mathbf{m}_{\mathrm{ext}}&=&0,\label{mbal}
\end{eqnarray}
where $\mathbf{F}(s,t)$ is the force the internal elastic stresses exert
through the cross section at $s$ on the portion of the rod with
arclength less then $s$. The external force per unit length
$\mathbf{f}_{\mathrm{ext}}$ and moment per unit length
$\mathbf{m}_{\mathrm{ext}}$ are measured per unit arclength. In an
alternative but equivalent formulation, the elastic force and
moment per unit length arise from variational derivatives of the
energy
\begin{equation}
E = \int\rm{d}s~\left({A\over2}\kappa^2 +
{C\over2}\left(\Omega-\Omega_0\right)^2-\Lambda(s,t)\right),
\label{energy}
\end{equation}
where $\kappa^2=|\partial^2\mathbf{r}/\partial
s^2|^2=\Omega_1^2+\Omega_2^2$ is the square of the curvature, and
$\Lambda$ is the Lagrange multiplier associated with the
constraint of prescribed length, $L=L_0\exp(\sigma
t)$~\cite{twirl,bistable_helices}.

The external force per unit length $\mathbf{f}_\mathrm{ext}$
consists of a viscous drag force per unit length and an artificial
short-ranged repulsive force that prevents self-crossing:
$\mathbf{f}_\mathrm{ext}=\mathbf{f}_\mathrm{visc}+\mathbf{f}_\mathrm{sel
f}$. The external moment per unit length is purely viscous:
$\mathbf{m}_\mathrm{ext}=\mathbf{m}_\mathrm{visc}$. The repulsive
force takes the form
\begin{equation}
\mathbf{f}_\mathrm{self}(s,t)=\int_{|s-s^\prime|>\delta}{\beta({\bf
r}(s)-
{\bf r}(s^\prime))\over|{\bf r}(s)-{\bf r}(s^\prime)|^{n}}~{\rm
d}s^\prime,
\label{fself}
\end{equation}
where $\delta$ is a short-distance cutoff, $n=14$ simulates the
repulsive part of a Leonard-Jones potential, and $\beta$=180 is
sufficient to keep the filament from self crossing. The viscous
force per unit length $\mathbf{f}_\mathrm{visc}$ depends on the
velocity field of the ambient fluid, which in turn is coupled to
the motion of the filament. For simplicity, we do not solve the
full hydrodynamic problem, but instead use resistive force theory
for slender bodies~\cite{kr}. Resistive force theory amounts to
the leading terms in an expansion in the aspect ratio $a/L$ of
slender-body theory, which has a nonlocal relation between force
and velocity due to incompressibility. To leading order, the
nonlocality can be neglected, leading to the local drag law,
\begin{equation}
\mathbf{f}_{\mathrm{visc}}=-\zeta_\perp(\mathbf{v}-\mathbf{\hat e}_3\
\mathbf{\hat e}_3
\cdot\mathbf{v})-\zeta_\parallel\mathbf{\hat e}_3\ \mathbf{\hat e}_3
\cdot\mathbf{v},
\label{resistive_force}
\end{equation}
Likewise, we take the viscous moment to be proportional to the
tangential component of the angular velocity of the material
frames,
\begin{equation}
\mathbf{m}_{\mathrm{visc}}=-\zeta_\mathrm{R}\mathbf{\hat e}_3\
\mathbf{\hat e}_3\cdot\bom.
\label{resistive_moment}
\end{equation}
The friction coefficients in Eqs.~(\ref{resistive_force}) and
(\ref{resistive_moment}) are
\begin{equation}
\zeta_\perp={4\pi\eta\over\log({L\over2a})+{1\over2}},\quad\quad\quad
\zeta_\parallel={2\pi\eta\over\log({L\over2a})-{1\over2}},\quad\quad\quad
\zeta_\mathrm{R}=4\pi\eta a^2,
\label{zetas}
\end{equation}
where $\eta$ is the viscosity of the ambient fluid, $L$ is the
total contour length of the rod, and $a$ is the rod radius. For
simplicity, we disregard the anisotropy and define
$\zeta\equiv\zeta_\perp=\zeta_\parallel=4\pi\eta/\log(L_0/2a)$,
where $L_0=L(t=0)$. Note that we have kept only the leading order
terms in the logarithm of the initial aspect ratio.
%
%
Thus, $\mathbf{f}_\mathrm{ext}=-\zeta\mathbf{v}$.  These
assumptions lead to qualitative differences with the exact theory.
For example, the neglect of hydrodynamic interactions implicit in
the local drag approximation of Eqs.~(\ref{resistive_force}) and
(\ref{resistive_moment}) will affect the time-dependence of the
shape of the rod near self-contact points just before contact (see
{\it e.g.}~\cite{shelley}). Also, the assumption of isotropy
implies that the center of mass of a deforming closed loop remains
fixed~\cite{becker_koehler_stone}, whereas in the exact theory the
center of mass can move. These limitations of the simplified
hydrodynamic theory do not prevent it from capturing the essential
physics of the phenomena we wish to study, such as the onset of
buckling instabilities and the subsequent evolution of complex
shapes.

It is convenient for the numerical calculations of
section~\ref{numeric} to write the equations of motion in terms of
position $\mathbf{r}$ and twist strain $\Omega_3$.  To this end,
substitute the constitutive relation (\ref{khirch_const}) into the
moment balance equations (\ref{mbal}) to find the force
$\mathbf{F}$ on a cross-section
\begin{equation}
\mathbf{F}=-A{\partial^3\mathbf{r}\over\partial
s^3}+C(\Omega_3-\tau_0){\partial\mathbf{r}
\over\partial s}\times{\partial^2\mathbf{r}\over\partial
s^2}-\Lambda{\partial\mathbf{r}
\over\partial s},
\label{Feqn}
\end{equation}
and the balance of the tangential components of the moment per unit
length,
\begin{equation}
C{\partial\Omega_3\over\partial s}=\zeta_\mathrm{R}\omega_3.
\label{tan_mom}
\end{equation}

The unknown function $\Lambda(s,t)$ occurs in Eq.~(\ref{Feqn})
because the moment balance equation (\ref{mbal}) does not
determine the tangential component of $\mathbf{F}$.  Combining
Eq.~(\ref{Feqn}) with the force balance equation (\ref{fbal}) and
the expressions for $\mathbf{f}_\mathrm{ext}$ yields
\begin{equation}
\zeta\left. {\partial\mathbf{r}\over\partial
t}\right|_{s_0}=-A{\partial^4\mathbf{r}\over
\partial s^4}+C{\partial\over\partial
s}\left[(\Omega_3-\tau_0){\partial\mathbf{r}\over
\partial s}\times{\partial^2\mathbf{r}\over\partial
s^2}\right]-{\partial\over\partial s}
\left[\Lambda{\partial\mathbf{r}\over\partial s}\right] +
\mathbf{f}_\mathrm{self}.
\label{r-eqn}
\end{equation}
Equation~(\ref{r-eqn}) has the same form as the corresponding
equation for the overdamped dynamics of an inextensible rod, but
in Eq.~(\ref{r-eqn}) the $s$-domain (length) depends on $t$. To
determine $\Lambda$, evaluate $\mathbf{\hat e}_3\cdot\partial
\mathbf{v}/\partial s= \sigma$ using Eq.~(\ref{r-eqn}):
\begin{eqnarray}
{\partial^2\Lambda\over\partial s^2}-{\partial^2\mathbf{r}\over\partial
s^2}\cdot
{\partial^2\mathbf{r}\over\partial s^2}\Lambda&=&-A{\partial
\mathbf{r}\over\partial s}\cdot
{\partial^5 \mathbf{r}\over\partial s^5}-\zeta\sigma+
{\partial \mathbf{r}\over\partial
s}\cdot\mathbf{f}_\mathrm{self}\nonumber\\
&+&C(\Omega_3-\tau_0){\partial \mathbf{r}\over\partial
s}\cdot{\partial^2\mathbf{r}\over
\partial s^2}\times{\partial^3\mathbf{r}\over\partial s^3}.
\label{Lambdaeqn}
\end{eqnarray}
One can show that the function $\Lambda$ in Eq.~(\ref{Lambdaeqn})
is identical to the Lagrange multiplier function of
Eq.~(\ref{energy}).

To complete the determination of the dynamical equations, the
torque balance (\ref{tan_mom}) and compatibility relation
(\ref{comp3}) yield
\begin{equation}
\left. {\partial\Omega_3\over\partial
t}\right|_{s_0}=D{\partial^2\Omega_3\over\partial s^2}
-\sigma\Omega_3+{\partial\mathbf{r}\over\partial
s}\times{\partial^2\mathbf{r}\over\partial
s^2}\cdot{\partial\mathbf{r}\over\partial t},
\label{twistdyn}
\end{equation}
where the twist diffusion constant $D\equiv C/\zeta_\mathrm{R}$.
Since resistive force theory includes the leading order terms in
the expansion in $a/L$ of the hydrodynamic drag force and torque,
our equations are asymptotically consistent.

The boundary conditions which accompany
Eqs.~(\ref{r-eqn}--\ref{twistdyn}) depend on the situation. For a
filament with free ends, the appropriate conditions are
$\mathbf{M}=0$ and $\mathbf{F}=0$.  For a closed loop, the
variables $\mathbf{r}$, $\bOm$, $\bom$, $\mathbf{F}$, and
$\mathbf{M}$ must be periodic in $s_0$ with period $2\pi R_0$.

\subsection{Change of Basis}
\label{change_of_basis}
We close this section by returning to the claim of
section~\ref{frame_growth_model}
that the spontaneous twist $\tau_0$ does not affect the shape of
an inextensible rod with circular cross-section and vanishing
spontaneous curvature $\kappa_0=0$. Consider such a rod with
$\mathbf{\Omega}=\tau_0\mathbf{\hat e}_3$ in the stress-free
state. We can eliminate $\tau_0$ from the problem using the
freedom to redefine the material frame. If
\begin{eqnarray}
\mathbf{\hat e}^\prime_1&=&\mathbf{\hat e}_1\cos\phi
-\mathbf{\hat e}_2\sin\phi,\label{erot1}\\
\mathbf{\hat e}^\prime_2&=&\mathbf{\hat e}_1\sin\phi
+\mathbf{\hat e}_2\cos\phi,
\label{erot2}
\end{eqnarray}
then
\begin{eqnarray}
\Omega^\prime_1&=&\Omega_1\cos\phi-\Omega_2\sin\phi\label{O1}\\
\Omega^\prime_2&=&-\Omega_1\sin\phi+\Omega_2\cos\phi\label{O2}\\
\Omega^\prime_3&=&\Omega_3+{\partial\phi\over\partial s}.\label{O3}
\end{eqnarray}
and
\begin{eqnarray}
\omega^\prime_1&=&\omega_1\cos\phi-\omega_2\sin\phi\label{o1}\\
\omega^\prime_2&=&-\omega_1\sin\phi+\omega_2\cos\phi\label{o2}\\
\omega^\prime_3&=&\omega_3+\left. {\partial\phi\over\partial
t}\right|_{s_0}.\label{o3}
\end{eqnarray}
Under this change of basis,
$\Omega^\prime_1\mathbf{\hat e}^\prime_1+
\Omega^\prime_2\mathbf{\hat e}^\prime_2=\Omega_1\mathbf{\hat e}_1
+\Omega_2\mathbf{\hat e}_2$ for any $\phi$. Choosing $\phi=-s\tau_0$
fixes
$\Omega^\prime_3=0$ in the stress-free state.  Thus,
\begin{equation}
\mathbf{M}=A(\Omega^\prime_1\mathbf{\hat e}^\prime_1+
\Omega^\prime_2\mathbf{\hat e}^\prime_2)+C\Omega^\prime_3\mathbf{\hat
e}^\prime_3;
\label{newconst}
\end{equation}
the parameter $\tau_0$ has been eliminated from the constitutive
relation. Note that our argument up to this point holds for a
growing rod as well.

Now consider the effect of the transformation
(\ref{erot1}--\ref{erot2}) on the compatibility relations. Once
again, even in the presence of exponential growth, the
compatibility equations take the same form, {\it e.g.}
\begin{equation}
\left. {\partial\Omega^\prime_3\over\partial t}
\right|_{s_0}={\partial\omega^\prime_3\over\partial
s}-\sigma\Omega^\prime_3
+\Omega^\prime_1\omega^\prime_2-\Omega^\prime_2\omega^\prime_1.
\label{comp3prime}
\end{equation}
For an inextensible rod, $\omega^\prime_3=\omega_3$, since $s=s_0$
if $\sigma=0$. Therefore, the change of basis
(\ref{erot1}--\ref{erot2}) does not affect the rotational drag or
translational drag equations, and we conclude that $\tau_0$ is not
a physical parameter for an inextensible rod with circular cross-section
and no spontaneous curvature.  However, we expect the
opposite conclusion for a growing rod, since we saw in
section~\ref{frame_growth_model}
that $\tau_0$ has physical meaning. In fact, once $\tau_0$ is
eliminated from the constitutive relation using
Eqs.~(\ref{erot1}--\ref{erot2}), a new $\tau_0$-dependent term
appears in the torque balance equation (\ref{tan_mom}):
\begin{equation}
C{\partial\Omega^\prime_3\over\partial s}=\zeta_\mathrm{R}
(\omega^\prime_3+\sigma s\tau_0).
\label{tan_mom_prime}
\end{equation}
Therefore, $\tau_0$ cannot be eliminated from the equations of
motion for this mode of growth.

\section{The Growing Elastic Loop}
\label{grow_elas_loop}

In this section we treat the problem of a growing elastic loop
with preferred twist $\tau_0$. As described in the introduction,
this example is motivated by the growth of {\it B.~subtilis}
filaments from a spore. Sometimes the ends of the growing filament
stick to the spore coat, leading to a closed
loop~\cite{mendelson76}. The filament lengthens, depleting the
twist, but the closed geometry prevents the rotation of
cross-sections normally seen in unconstrained filaments. Thus,
twist stress builds up, and the filament eventually writhes and
coils. Although the model does not address the writhing and
coiling of unconstrained filaments, it displays some of the
features exhibited by {\it B.~subtilis} loops.

\subsection{Buckling and Writhing Instabilities}
\label{buck-writhe} To study the stability of an exponentially
growing circular loop with preferred twist $\tau_0$, we begin with
the unperturbed solution.  The unperturbed loop lies in the $z=0$
plane and has radius $R=R_0\exp(\sigma t)$. Since each material
point on the filament moves radially outward with fixed $z$ in
unperturbed growth, the angular velocity vanishes, $\bom^{(0)}=0$,
where we use the superscript, $(0)$, to denote the unperturbed
value. The bending part of the energy (\ref{energy}) decreases
exponentially as the loop grows because $\kappa^{(0)}=1/R$.
However, the twist energy density {\it increases} because the
closed geometry prevents the cross-sections from rotating with the
rate $\sigma s\tau_0$ required to attain the twist state of zero
energy. If we assume for simplicity that
$\Omega_3^{(0)}(s,t=0)=\tau_0$, then the exponentially increasing
length leads to an exponentially decreasing twist density,
$\Omega_3^{(0)}(s,t)=\tau_0\exp(-\sigma t)$. To summarize, the
moment on a cross-section takes the form
\begin{equation}
{\mathbf M}^{(0)}={A\over R_0}\e^{-\sigma t}\zhat+C\tau_0
\bigg(\e^{-\sigma t}-1\bigg)\phihat. \label{Mzero}
\end{equation}
Note that it is much more convenient to express ${\mathbf M}$ in
terms of the cylindrical coordinate unit vectors
$\{\zhat,\rhohat,\phihat\}$ instead of the material frame vectors
$\{\ehat_1^{(0)},\ehat_2^{(0)},\ehat_3^{(0)}\}$, since
$\ehat_1^{(0)}$ and $\ehat_2^{(0)}$ continuously rotate about
$\ehat_3^{(0)}$ as $s$ increases.  A choice of frame which does not
rotate about
the tangent vector as arclength increases
is a {\it natural} frame~\cite{bishop}. We will
use the natural frame extensively in the linear stability
analysis of section~\ref{lin_stab_anal}.

Since the filament is simultaneously bent and twisted, moment
balance (\ref{mbal}) implies a component of force in the
$z$-direction, whereas force balance (\ref{fbal}) leads to a
tangential component proportional to the growth rate $\sigma$:
\begin{equation}
{\mathbf F}^{(0)}=-\zeta\sigma R^2\phihat+{C\tau_0\over R}
\bigg(\e^{-\sigma t}-1\bigg)\zhat. \label{Fzero}
\end{equation}
The tangential force on the cross-section is compressive and grows
exponentially in time, eventually leading to an Euler-like
buckling instability when $\zeta\sigma R^2\approx A/R^2$, or
$R=R_1\approx[A/(\zeta\sigma)]^{1/4}$ ({\it cf.}
\cite{shelley,shelleyUeda,drasdo}). The Euler buckling time scales
as $t_\mathrm{E}\approx\sigma^{-1}\log[A/(\zeta\sigma R_0^4)]$
(when $A/(\zeta\sigma R_0^4)<1$, the loop begins to buckle at
$t=0$). In section~\ref{lin_stab_anal}, we will refine this
estimate using our linear stability calculation and see that the
appropriate buckling time at small $\sigma$ actually scales as
$\sigma^{-2}$, since in this regime, a growing perturbation does
not become noticeable until long after the perturbation begins to
grow. However, our numerical calculations of section~\ref{numeric}
reveal that the correct picture is even more complicated: for
sufficiently small $\tau_0R_0$, a secondary instability arises and
grows before the linear instability has significant amplitude. In
addition to the Euler buckling, there is also a writhing
instability. The magnitude of the twist moment increases as
$|C\tau_0(1-\exp(-\sigma t))|$ as the filament lengthens, leading
to a writhing instability (like that of a twisted
ring~\cite{michell,basset,zajac}) when $C\tau_0(1-\exp(-\sigma
t))\approx A/(2\pi R)$, or
$t_\mathrm{W}\approx\sigma^{-1}\log[1+1/(\tau_0 R_0)]$, (assuming
$A\approx C$).  Note that the critical time for writhing is
$t_\mathrm{W}\approx\sigma^{-1}\log[1/(R_0\tau_0)]$ for
$R_0\tau_0\ll1$, and $t_\mathrm{W}\approx1/(\sigma R_0\tau_0)$ for
$R_0\tau_0\gg1$. The sense of rotation of the cross-sections of
unconstrained filaments determines the handedness of the coils
that form after the instability of the ring:  positive $\tau_0$
(counter-clockwise rotation when viewed from the direction of
increasing $s$) leads to right-handed plectonemic braids. Note
that drag is the ultimate cause for the Euler buckling
instability; once growth ceases, the buckled filament relaxes back
to the unperturbed circular shape. Since the writhing instability
arises not from drag but instead from the frustration of
growth-induced twist stress, the braided post-instability shape
remains after growth ceases. For small enough $\tau_0R_0$, we will
see in section~\ref{numeric} that the small intrinsic twist biases
the Euler buckling, leading to solenoidal shapes, which relax to
plectonemes if growth halts after a sufficiently long time.
Observations of the growing fibers suggest that writhing is the
dominant mechanism in the instability of a closed
loop~\cite{mendelson76}. Using $A=10^{-12}$ erg-cm \cite{walk},
$\sigma=2\times10^{-4}$ s$^{-1}$, $L=10^{-3}$ cm,
$a=3\times10^{-5}$ cm, and $\zeta=\zeta_{\perp}\approx 10^{-1}$
erg-s/cm$^3$ (see Eqn.~(\ref{zetas})) leads to
$[A/(\zeta\sigma)]^{1/4}\approx150$ $\mu$m, which is much larger
than the observed critical radius and implies that growth-induced
Euler buckling is not important. Thus, we can use the critical
radius of the twisted ring to estimate $\tau_0$; assuming $\tau_0$
is comparable to or smaller than $1/R_0$, where $R_0$ is the
initial radius, and using $R_2\approx2\times10^{-4}$ cm as the
experimentally observed critical radius~\cite{mendelson76}, we
find $\tau_0\approx10^{4}$ cm$^{-1}$.  This twist rate corresponds
to a few turns per cell. It is intriguing to note that the
corresponding length scale is close to the pitch of helical
filaments of mbl, a recently discovered protein which resides near
the cell wall and plays a role in maintaining the shape of {\it B.
subtilis} cells~\cite{errington}.

\subsection{The Natural Frame}
\label{natframe}

Before studying the evolution of small perturbations of the
growing circular shape, we consider the choice of representation.
Since only the shape and twist are of interest, it is convenient
to use an intrinsic representation, such as the material frame.
However, we saw in the solution of the unperturbed problem of
section~\ref{buck-writhe} that the natural frame leads to further
simplification. The advantages of the natural frame over the
material frame are even greater for the linear stability analysis,
and more generally for the full nonlinear problem
~\cite{bistable_helices,goldstein_etal1998}.

In a natural orthonormal frame $\{\nhat_1,\nhat_2,\ehat_3\}$ the
instantaneous rate of rotation of $\nhat_1$ about $\ehat_3$ is
zero~\cite{bistable_helices,bishop,goldstein_etal1998}:
\begin{equation}
\nhat_1\cdot{\partial\nhat_2\over\partial s}=0. \label{nat_frame1}
\end{equation}
A rotation of $\{\nhat_1,\nhat_2\}$ about $\ehat_3$
by uniform (arclength-independent) angle leaves the
condition~(\ref{nat_frame1}) invariant; every space curve has a family of
natural frames, the members of which are related to each other by
rotation through a uniform angle. To construct a natural frame from the
material
frame $\{\ehat_1,\ehat_2,\ehat_3\}$, rotate the material frame at
$s$ by minus the accumulated rotation angle
$\vartheta=\int_0^s{\mathrm d}s^\prime\Omega_3$:
\begin{eqnarray}
\nhat_1&=&\cos\vartheta\ehat_1-\sin\vartheta\ehat_2\label{rot_frame1}\\
\nhat_2&=&\sin\vartheta\ehat_1+\cos\vartheta\ehat_2.
\label{rot_frame2}
\end{eqnarray}
The natural frame is nonlocal in the sense that deformations of
the filament centerline in the region $s^\prime<s$ affect the
natural frame at $s$.

Our formulas can be further simplified with complex notation. For
example, if $\bep\equiv\nhat_1+\mathrm{i}\nhat_2$, then
Eqs.~(\ref{rot_frame1}--\ref{rot_frame2}) become
\begin{equation}
\bep=(\ehat_1+\mathrm{i}\ehat_2)\e^{\mathrm{i}\vartheta}.
\label{ep_def}
\end{equation}
The vector $\bep$ is a complex normal vector. Note that
$\bep\cdot\bep=0$; therefore, $\bep\cdot\partial\bep/\partial
s=0$.  Furthermore, the defining property
$\nhat_1\cdot\partial\nhat_2/\partial s=0$ implies
$\bep^*\cdot\partial\bep/\partial s=0$. Thus,
$\partial\bep/\partial s$ is proportional to $\ehat_3$. In analogy
with the Frenet-Serret equation $\partial\nhat/\partial
s=-\kappa\that$, we define the complex curvature $\Psi$ via
\begin{equation}
{\partial\bep\over\partial s}=-\Psi\ehat_3, \label{d-ep}
\end{equation}
where
\begin{eqnarray}
\Psi&=&-\ehat_3\cdot{\partial\bep\over\partial s}
=\bep\cdot{\partial\ehat_3\over\partial s}\label{e3-dot-ep_s}\\
&=&(-\mathrm{i}\Omega_1+\Omega_2)\e^{\mathrm{i}\vartheta}.
\label{psidef2}
\end{eqnarray}
Also, the rate of change along $s$ of the unit tangent vector
is the complex curvature
times the complex normal vectors,
\begin{equation}
{\partial\ehat_3\over\partial
s}={1\over2}\bigg(\Psi\bep^*+\Psi^*\bep\bigg). \label{d-e3}
\end{equation}
Note that rotation of a natural frame about $\ehat_3$ by a uniform
angle leads to a constant shift in the phase of $\Psi$. For
example, the natural frame arising from the
construction~(\ref{rot_frame1}--\ref{rot_frame2}) applied to the
Frenet-Serret normal and binormal has
\begin{equation}
\bep=(\nhat+\mathrm{i}\bhat)\exp\Bigg(\mathrm{i}\int\tau\mathrm{d}s\Bigg),
\label{ep-def2}
\end{equation}
since torsion is the rate at which $\nhat$ and $\bhat$ twist around
the tangent vector.  The corresponding complex curvature is
\begin{equation}
\Psi_{\mathrm{FS}}=\kappa\exp\Bigg(\mathrm{i}\int\tau\mathrm{d}s\Bigg);
\label{Psi-def2}
\end{equation}
the ratio of $\Psi$ and $\Psi_{\mathrm{FS}}$ is a constant phase.

To complete the specification of the kinematics of the natural
frame, we introduce the complex angular velocity describing
the rate of change with time of the unit tangent vector:
\begin{eqnarray}
\Pi&\equiv&\bep\cdot{\partial\ehat_3\over\partial t}\\
&=&\bigg(-\mathrm{i}\omega_1+\omega_2\bigg)\e^{\mathrm{i}\vartheta}.\label{Pidef}
\end{eqnarray}

In terms of the natural frame variables, the compatibility
relations (\ref{comp1}--\ref{comp3}) become
\begin{eqnarray}
\left. {\partial\Psi\over\partial
t}\right|_{s_0}&=&{\partial\Pi\over\partial
s}-\sigma\Psi-\mathrm{i}\Psi\omega_3(0)+\mathrm{i}\Psi\int_0^s
{\rm d}s^\prime\mathrm{Im}\bigg(\Psi^*\Pi\bigg)\label{Psi-comp}\\
\left. {\partial\Omega_3\over\partial
t}\right|_{s_0}&=&{\partial\omega_3\over\partial
s}-\sigma\Omega_3+\mathrm{Im}\bigg(\Psi^*\Pi\bigg)\label{Om-comp}.
\end{eqnarray}
The integral in Eq.~(\ref{Psi-comp}) reflects the
nonlocality of the natural frame, and arises from the temporal rate of
change of $\vartheta$:
\begin{eqnarray}
\left. {\partial\vartheta\over\partial t}\right|_{s_0}&=&
\left. {\partial\over\partial t}\right|_{s_0}
\int_0^{s_0\e^{\sigma t}}\Omega_3(s^\prime,t)\,
\mathrm{d}s^\prime\label{theta_t1}\\
&=&\left. {\partial\over\partial t}\right|_{s_0}
\int_0^{s_0}\Omega_3(s_0^\prime\e^{\sigma t},t)\e^{\sigma t}\,
\mathrm{d}s_0^\prime\label{theta_t2}\\
&=&\int_0^s\bigg[{\partial\Omega_3\over\partial
t}+\sigma\Omega_3\bigg]\,\mathrm{d}s^\prime;
\label{theta_t3}
\end{eqnarray}
Eq.~(\ref{Psi-comp}) follows from Eqs.~(\ref{theta_t3}),
(\ref{comp1}--\ref{comp3}),
and the identity
$\Omega_1\omega_2-\Omega_2\omega_1=\mathrm{Im}(\Psi^*\Pi).$

In the natural frame variables, the moment $\mathbf{M}$ obeys
\begin{equation}
\mathbf{M}=A{\mathrm{i}\over2}\bigg(\Psi\bep^*
-\Psi^*\bep\bigg)+C\bigg(\Omega_3-\tau_0\bigg)\ehat_3.
\label{M-nat}
\end{equation}
Just as in section~\ref{constit},
moment balance~(\ref{mbal}) determines the perpendicular component
of the force on a cross-section
\begin{eqnarray}
\mathbf{F}&=& F_\parallel\ehat_3+{1\over2}\bigg(\bep^*F_\perp+\bep
F^*_\perp\bigg);\label{F-nat}\\
F_\perp&=&-A{\partial\Psi\over\partial
s}+\mathrm{i}C\Psi\bigg(\Omega_3-\tau_0\bigg). \label{Fperp-nat}
\end{eqnarray}
Recall that $F_\parallel$ is not determined by moment balance since
only the perpendicular components of $\mathbf{F}$ enter Eq.~(\ref{mbal});
$F_\parallel$ is determined by the condition
$\ehat_3\cdot\partial\mathbf{v}/\partial
s=\sigma$.

Defining the tangential and perpendicular
components $f_\parallel$ and $f_\perp$ of the force per unit length
\begin{equation}
{\partial\mathbf{F}\over\partial
s}=\ehat_3f_\parallel+{1\over2}\bigg(\bep^*f_\perp+\bep
f_\perp^*\bigg),\label{Fs-nat}
\end{equation}
it follows from Eqs.~(\ref{d-ep}), (\ref{d-e3}), (\ref{F-nat}),
and (\ref{Fperp-nat}) that
\begin{equation}
f_\parallel={\partial F_\parallel\over\partial
s}+{A\over2}\bigg(\Psi\Psi_s^*+\Psi_s\Psi^*\bigg),\label{f-par-nat}
\end{equation}
and
\begin{equation}
f_\perp=F_\parallel\Psi-A{\partial^2\Psi\over\partial
s^2}+\mathrm{i}C{\partial\over\partial
s}\bigg[\Psi\bigg(\Omega_3-\tau_0\bigg)\bigg].\label{f-perp-nat}
\end{equation}
Note $f_\parallel=\partial(F_\parallel+A|\Psi|^2/2)/\partial
s=-\partial\Lambda/\partial s$; in terms of $f_\parallel$ and $f_\perp$
the condition $\ehat_3\cdot\partial\mathbf{v}/\partial
s=\sigma$ becomes
\begin{equation}
\zeta\sigma={\partial f_\parallel\over\partial s}
-{1\over2}f_\perp\Psi^*-{1\over2}f_\perp^*\Psi.\label{sigma-eqn}
\end{equation}
Finally, the force per unit length determines
the complex angular velocity through the relations
$\zeta\mathbf{v}=\partial\mathbf{F}/\partial s$ and
$\bep\cdot\partial\mathbf{v}/\partial s=\Pi$, or
\begin{equation}
\zeta\Pi={\partial f_\perp\over\partial s}+\Psi f_\parallel.\label{Pifperp}
\end{equation}
Likewise, the  tangential component of the angular velocity is
given by Eq.~(\ref{tan_mom}). In summary, the equations of motion
for the filament in the natural frame are
Eqs.~(\ref{Psi-comp}--\ref{Om-comp}),
(\ref{f-par-nat}--\ref{Pifperp}), and (\ref{tan_mom}).

\subsection{Linear Stability Analysis}
\label{lin_stab_anal}

We now return to the stability analysis of a growing circular
ring. We write the shape as
$\mathbf{r}=(R+r^{(1)})\rhohat+z^{(1)}\zhat$ and work to first
order in $r^{(1)}$ and $z^{(1)}$. It is convenient to find the
equations of motion for $\Psi^{(1)}$ and $\Omega_3^{(1)}$ first,
and then express these equations in terms of $r^{(1)}$ and
$z^{(1)}$. To this end, we choose
$\{\nhat_1^{(0)},\nhat_2^{(0)}\}=\{-\rhohat,\zhat\}$ to make
$\Psi^{(0)}=1/R$. Expanding~(\ref{Psi-def2}) to first order leads
to
\begin{equation}
\Psi=\kappa^{(0)}+\kappa^{(1)}+\mathrm{i}\kappa^{(0)}\int\tau^{(1)}\mathrm{d}
s,
\label{psi-expand}
\end{equation}
since the unperturbed loop has zero torsion, $\tau^{(0)}=0$. To
express $\kappa$ and $\tau$ in terms of $r^{(1)}$ and $z^{(1)}$,
use the Frenet-Serret equations~(\ref{F-S1}--\ref{F-S2}) and
$s=R\phi$ (to leading order) to find
\begin{equation}
\kappa\approx{1\over R}-{1\over R^2}(r^{(1)}+r^{(1)}_{\phi\phi})
\label{kappa-approx}
\end{equation}
and
\begin{equation}
\tau\approx\tau^{(1)}={1\over R^2}(z^{(1)}_\phi+z^{(1)}_{\phi\phi\phi}).
\label{torsion-approx}
\end{equation}
Thus, $\Psi=1/R+\xi+\mathrm{i}\eta$, with
\begin{eqnarray}
\xi&=&-{1\over R^2}(r^{(1)}+r^{(1)}_{\phi\phi})\label{xi-eqn}\\
\eta&=&{1\over R^2}(z^{(1)}+z^{(1)}_{\phi\phi})\label{eta-eqn}.
\end{eqnarray}
As described in section~\ref{buck-writhe}, the
unperturbed angular velocities vanish: $\Pi^{(0)}=0$,
$\omega_3^{(0)}=0$; furthermore,
Eq.~(\ref{Fzero}) together with (\ref{f-par-nat}--\ref{f-perp-nat}) imply
$f^{(0)}_\perp=F^{(0)}_\parallel/R=-\zeta\sigma R$ and
$f^{(0)}_\parallel=0$.
Expanding the equations of motion to first order, we find
\begin{eqnarray}
\left. {\partial\Psi^{(1)}\over\partial
t}\right|_{s_0}&=&{\partial\Pi^{(1)}\over\partial
s}-\sigma\Psi^{(1)}-\mathrm{i}\Psi^{(0)}\omega^{(1)}_3(0)+\mathrm{i}\Psi^{(0)}\int_0^s
{\rm
d}s^\prime\mathrm{Im}\bigg(\Psi^{(0)*}\Pi^{(1)}\bigg)\label{Psi-comp-pert}\\
\left. {\partial\Omega^{(1)}_3\over\partial
t}\right|_{s_0}&=&D{\partial^2\Omega^{(1)}_3\over\partial
s^2}-\sigma\Omega^{(1)}_3+\mathrm{Im}\bigg(\Psi^{(0)*}\Pi^{(1)}\bigg)\label{Om-comp-pert}\\
0&=&{\partial f^{(1)}_\parallel\over\partial s}
-\mathrm{Re}\bigg(f^{(0)}_\perp\Psi^{(1)*}\bigg)
-\mathrm{Re}\bigg(f^{(1)}_\perp\Psi^{(0)*}\bigg)\label{sigma-zeta-pert}\\
\zeta\Pi^{(1)}&=&{\partial f^{(1)}_\perp\over\partial s}+
\Psi^{(0)} f^{(1)}_\parallel,\label{Pifperp-pert}
\end{eqnarray}
where $f^{(1)}_\parallel$ and $f^{(1)}_\perp$ are determined by
expanding (\ref{f-par-nat}--\ref{f-perp-nat}) to first order.
Inspection of (\ref{Psi-comp-pert}--\ref{Pifperp-pert}) and
(\ref{f-par-nat}--\ref{f-perp-nat}) reveals that $r^{(1)}=\hat
r^{(1)}\cos(n\phi)$, $z^{(1)}=\hat z^{(1)}\sin(n\phi)$,
$F^{(1)}_\parallel=\hat F_\parallel\cos(n\phi)$, and
$\Omega_3^{(1)}=\hat\Omega_3\cos(n\phi)$, with $n$ a positive
integer. This choice of origin for $s$ and Eq.~(\ref{tan_mom})
imply $\hat\omega^{(1)}_3(0)=0$. The perturbation $\hat
r^{(1)}\cos\phi$ corresponds to a translation of the ring in the
$z=0$ plane; likewise, the perturbation $\hat z^{(1)}\sin\phi$
corresponds to a rotation of the ring about an axis in the $z=0$
plane. Thus, the $n=1$ perturbations are rigid motions, leading to
no change in curvature: $\xi=\eta=0$ (see
Eqs.~(\ref{xi-eqn}--\ref{eta-eqn})). Inserting the perturbations
into~(\ref{Psi-comp-pert},\ref{Om-comp-pert},\ref{Pifperp-pert})
and using~(\ref{sigma-zeta-pert}) to eliminate $\hat F_\parallel$
yields a linear system of differential equations for $\hat
r^{(1)}$, $\hat z^{(1)}$, and $\hat\Omega_3$:
\begin{eqnarray}
{\hat r}^{(1)}_t&=&\Bigg[\sigma{n^4+3\over n^2+1}-{A\over\zeta
R^4}{n^2(n^2-1)^2\over n^2+1}\Bigg]\hat
r^{(1)}\nonumber\\&&-{C\over\zeta R^3}(\Omega_3^{(0)}
-\tau_0){n^3(n^2-1)\over n^2+1}\hat z^{(1)}\label{growth-eq1}\\
\hat z^{(1)}_t&=&-{C\over\zeta
R^3}(\Omega_3^{(0)}-\tau_0)n(n^2-1)\hat r^{(1)}+\Bigg[\sigma
n^2-{A\over\zeta
R^4}n^2(n^2-1)\Bigg]\hat z^{(1)}\nonumber\\
&&-{C\over\zeta R^2}n\hat\Omega_3\label{growth-eq2}\\
\hat\Omega_{3t}&=&{C\over\zeta
R^3}(\Omega_3^{(0)}-\tau_0)n^2(1-n^2){\hat r^{(1)}\over R^2}
+\Bigg[{An^2\over\zeta
R^4}-\sigma\Bigg]n(1-n^2){\hat z^{(1)}\over R^2}\nonumber\\
&&-\Bigg[\sigma+{Dn^2\over R^2}+{Cn^2\over\zeta
R^4}\Bigg]\hat\Omega_3.\label{Om-t}\label{growth-eq3}
\end{eqnarray}
Equations (\ref{growth-eq1}--\ref{growth-eq2}) hold for $n\neq1$,
and equation (\ref{growth-eq3}) holds for all integer $n\ge1$. For
$n=1$,  the shape drops out of Eq.~(\ref{growth-eq3}) since the
$n=1$ shape perturbations are rigid motions. Note that $\tau_0$
enters (\ref{growth-eq1}--\ref{growth-eq3}) in the combination
$\Omega^{(0)}_3-\tau_0=\tau_0(\exp(-\sigma t)-1)$; $\tau_0$ drops
out of the equations when $\sigma=0$, in accord with the general
arguments of section~\ref{change_of_basis}.

We can simplify the linear
system~(\ref{growth-eq1}--\ref{growth-eq3}) by exploiting the
large ratio of relaxation time scales for twisting and bending
modes:
\begin{equation} t_\mathrm{bend}\gg t_\mathrm{twist};
\label{t-regime}\end{equation} for $R\approx10$ $\mu$m and the
{\it B. subtilus} parameters of section~\ref{buck-writhe},
$t_\mathrm{bend}\equiv\zeta R^4/A\approx10^{-1}$~s and
$t_\mathrm{twist}\equiv
R^2/D=\zeta_\mathrm{R}R^2/C\approx10^{-4}$~s. As time passes and
the filament lengthens, $t_\mathrm{bend}$ and $t_\mathrm{twist}$
increase exponentially, but $t_\mathrm{bend}\gg t_\mathrm{twist}$
for all time. Therefore, twist perturbations $\hat\Omega$ relax
immediately, and (\ref{growth-eq1}--\ref{growth-eq3}) may be
simplified by setting $\hat\Omega$ to zero. Using the initial
radius for the bending relaxation time ($t_\mathrm{bend}=\zeta
R_0^4/A$) and assuming $C/A=1$ for simplicity,
(\ref{growth-eq1}--\ref{growth-eq2}) reduce to
\begin{equation}\dot\mathbf{q}=\tensor{L}\mathbf{q},\label{lineq}\end{equation}
where
\begin{equation}
\mathbf{q}=\left(\matrix{\hat r^{(1)}\cr\hat z^{(1)}}\right)
\end{equation}
and
\begin{eqnarray}
\tensor{L}=\left[\matrix{\sigma{n^4+3\over n^2+1}-{\e^{-4\sigma
t}\over t_\mathrm{bend}} {n^2(n^2-1)^2\over
n^2+1}&-{\tau_0R_0\e^{-3\sigma t}\over
t_\mathrm{bend}}(\e^{-\sigma t} -1){n^3(n^2-1)\over
n^2+1}\cr-{\tau_0R_0\e^{-3\sigma t}\over
t_\mathrm{bend}}(\e^{-\sigma t} -1)n(n^2-1)&\sigma
n^2-{\e^{-4\sigma t}\over t_\mathrm{bend}}
n^2(n^2-1)}\right].\label{linearop}
\end{eqnarray}
Since $\tensor{L}$ depends on time, the system (\ref{lineq}) is
non-autonomous, and classical modal analysis~\cite{drazinreid1981}
does not apply. Note also that $\tensor{L}$ is not a normal
operator: $[\tensor{L},\tensor{L}^\mathrm{T}]\ne0$. Therefore, the
eigenvectors of $\tensor{L}$ are not perpendicular, which in
general signals the possibility of transient or algebraic growth
of perturbations~\cite{schmidhenningson2001}. However,
$\tensor{L}$ is only weakly non-normal, since the eigenvectors are
almost perpendicular. A similar conclusion applies to the problem
of a ring with twist but no growth ($\sigma=0$) in a viscous fluid
({\it cf.}~\cite{zajac,klapper94}). Therefore, we do not expect
the phenomenon of transient growth of perturbations in our
problem. Nevertheless, the methods developed to study non-normal
linear problems are well-suited to non-autonomous problems such as
(\ref{lineq})~\cite{schmid2000}.

We will characterize the growth of perturbations by the
amplification of the magnitude of $\mathbf{q}(0)$. The optimal
amplification $G(t)$ is defined by maximizing this factor over all
initial conditions:
\begin{equation}
G(t)\equiv\max_{\mathbf{q}(0)}{\mathbf{q}(t)\cdot\mathbf{q}(t)
\over\mathbf{q}(0)\cdot\mathbf{q}(0)}.
\label{optimalgrowthdef}
\end{equation}
(A more physical choice for the optimal growth rate would be to
use the bending energy to second order in $\hat r^{(1)}$ and $\hat
z^{(1)}$ instead of $\mathbf{q}\cdot\mathbf{q}$; it turns out that
either choice yields essentially the same $G(t)$.) To compute
$G(t)$, recast Eqn.~(\ref{lineq}) as an equation for the
propagator matrix $\tensor{B}$:
\begin{equation}
\dot\tensor{B}=\tensor{L}\tensor{B},
\end{equation}
where $\tensor{B}(0)=\tensor{I}$ and
$\mathbf{q}(t)=\tensor{B}(t)\mathbf{q}(0)$. Thus, the optimal
growth factor is the Rayleigh quotient
\begin{equation}
G(t)=\max_{\mathbf{q}(0)}{\mathbf{q}(0)\cdot\tensor{B}^\mathrm{T}(t)
\tensor{B}(t)\mathbf{q}(0)\over \mathbf{q}(0)\cdot\mathbf{q}(0)},
\end{equation}
or $G(t)=\lambda_+$, where $\lambda_+$ is the largest eigenvalue
of $\tensor{B}^\mathrm{T}(t)\tensor{B}(t)$~\cite{strang1988}. When
$\tensor{L}$ is a $t$-independent normal matrix, then
$\lambda_+=\exp(2\Lambda_+t)$, where $\Lambda_+$ is the largest
eigenvalue of $\tensor{L}$.  Note that since $G(t)$ is maximized
at each $t$ over all initial conditions $\mathbf{q}(0)$, the
maximum amplitudes at two different times may correspond to two
different initial conditions $\mathbf{q}(0)$. We computed $G(t)$
by using standard Runge-Kutta techniques to solve for
$\tensor{B}(t)$~\cite{highamhigham2000}.
\begin{figure}
\includegraphics[height=4in]{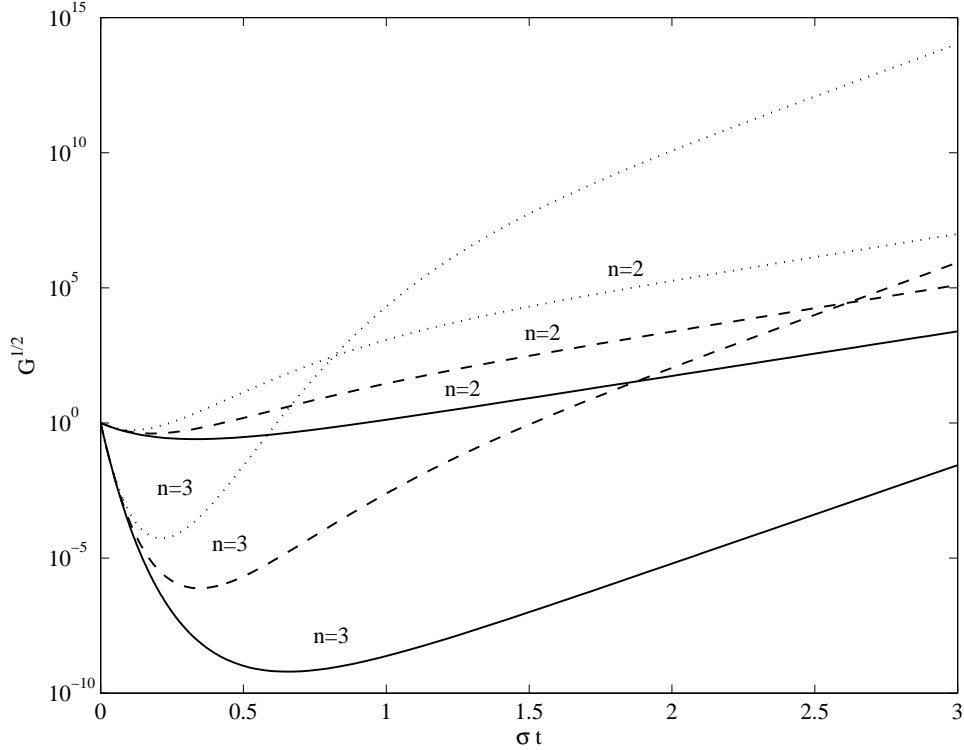}
\caption{Optimal growth factors for $\sigma t_\mathrm{bend}=0.5\
(<\sigma_\mathrm{c}t_\mathrm{bend})$. Solid line: $R_0\tau_0=0$.
Dashed line: $R_0\tau_0=5$. Dotted line: $R_0\tau_0=10$. Note that
the intermediate regime of rapid growth of $G(t)$ is most apparent
for $R_0\tau_0=10$, especially for $n=3$.} \label{taudep}
\end{figure}

Inspection of the diagonal components of $\tensor{L}$
(\ref{linearop}) reveals that for sufficiently rapid growth,
$\sigma>\sigma_\mathrm{c}\equiv
n^2(n^2-1)^2/[(n^2+3)t_\mathrm{bend}]$, the loop deforms away from
its circular shape as soon as it begins to grow. When the rate of
growth of the ring is sufficiently slow,
$\sigma<\sigma_\mathrm{c}$, bending stiffness stabilizes the
circular shape for $\sigma t\ll1$ and perturbations decay roughly
as $\exp(-n^4 t/t_\mathrm{bend})$. Thus, the growth factor $G$
decays extremely rapidly with increasing $n$ at early times. As
time passes, the loop lengthens and eventually buckles, with the
nature of the buckling dependent on the magnitude of $\tau_0$. For
$\tau_0R_0\ll1$, the distortion is the three-dimensional analog of
the in-plane Euler buckling studied by Shelley and
Ueda~\cite{shelleyUeda,shelley} and Drasdo~\cite{drasdo}. For
$\tau_0R_0\gg1$, the off-diagonal elements of $\tensor{L}$ are
large [see (\ref{linearop})], writhing dominates the nature of the
initial distortion, and $G(t)$ increases roughly as $\exp(2\tau_0
R_0 n^3t/t_\mathrm{bend})$ in the intermediate regime $\sigma
t<\approx 1$. In the late-time regime $\sigma t\gg1$,
$G(t)\propto\exp(2n^2\sigma t)$ for any value of $\tau_0R_0$ or
$t_\mathrm{b}$. These results are summarized in Fig.~\ref{taudep}.

\begin{figure}
\includegraphics[height=3.5in]{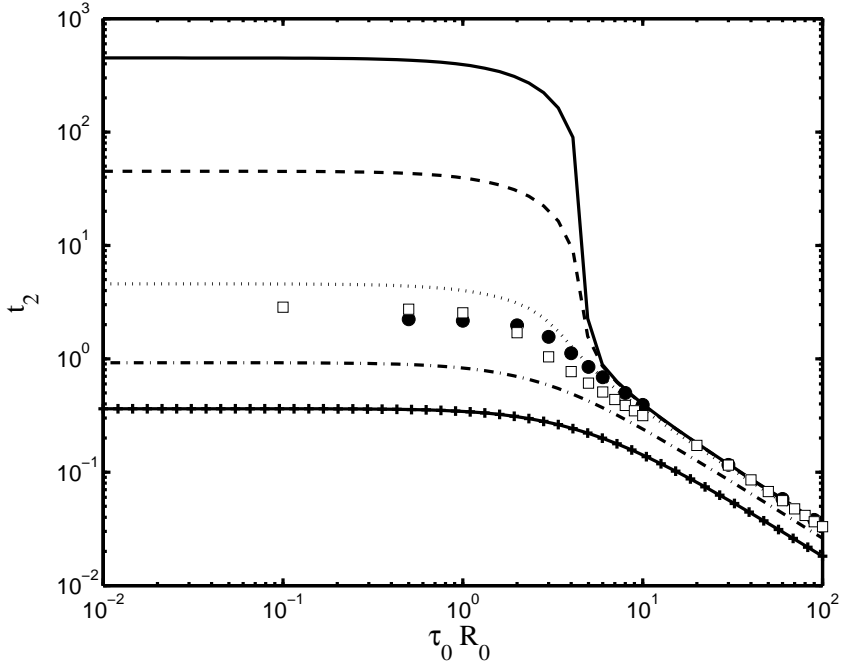}
\caption{Time $t_2$ for optimal perturbation to regain initial
amplitude as a function of $R_0\tau_0$ for $\sigma
t_\mathrm{bend}=0.001$ (solid line), $\sigma t_\mathrm{bend}=0.01$
(dashed line), $\sigma t_\mathrm{bend}=0.1$ (dotted line), $\sigma
t_\mathrm{bend}=0.5$ (dashed-dotted line), and $\sigma
t_\mathrm{bend}=1$ (solid line with small dots). The large filled
dots are the results of numerical simulations of the full
nonlinear equations with $\sigma t_\mathrm{bend}=0.01$ and the
open squares are with $\sigma t_\mathrm{bend}=0.1$; see
section~\ref{numeric}.} \label{tcrit}\end{figure}

There are two different times which may be chosen to represent the
time of the instability. The estimate for the buckling time of
section~\ref{buck-writhe} amounts to the time $t_1$ at which
$G(t)$ reaches its minimum value and starts to increase. However,
since the rate of decay of $G(t)$ in the stable period can be very
different from the rate of growth of $G(t)$ in the unstable
period, the time at which a perturbation becomes noticeable may be
significantly greater than $t_1$.  Thus, it is natural to define
the time for the onset of the instability to be the time $t_2$ at
which the amplitude of the optimal perturbation regains its
initial value: $G(t_2)=1$.  The graphs of Fig.~\ref{taudep}
suggest that $t_2$ and $t_1$ are comparable whenever $R_0\tau_0$
is large enough for the intermediate growth regime of rapid growth
discussed above to be present. However, if  $R_0\tau_0$ is small
enough that this intermediate regime is absent, then $t_2$ will be
much greater than $t_1$ when the extensional growth rate is slow,
$\sigma t_\mathrm{bend}\ll1$. Figure~\ref{tcrit} shows how
dramatic this difference can be. For $R_0\tau_0\ll1$ and $\sigma
t_\mathrm{bend}\ll1$,
\begin{equation}
t_2\approx{1\over\sigma^2t_\mathrm{bend}}{n^2(n^2-1)^2\over4(n^4+3)},
\label{t2approximation}
\end{equation}
whereas $t_1\propto1/\sigma$. When $R_0\tau_0\gg1$, both $t_1$ and
$t_2$ scale as $1/\sigma$ (see Fig.~\ref{tcrit}).  Thus, for small
$\sigma t_\mathrm{bend}$, there is a sharp transition in the onset
time $t_2$ as a function of $R_0\tau_0$. In section~\ref{numeric}
we will see how this prediction of the linear theory captures the
early-time dynamics for $R_0\tau_0\gg1$, but that nonlinearities
intervene before $t=t_2$ for $R_0\tau_0\ll1$.

The curves for the onset time $t_2$ of Fig.~\ref{tcrit} were
computed from the linearized equations (\ref{lineq}) using the
adiabatic theorem. If $\sigma t_\mathrm{bend}\ll1$, and if
$\tensor{L}$ were normal, then the adiabatic
theorem~\cite{messiahII} would imply that
\begin{equation}
\tensor{B}\approx\mathbf{v}_+\mathbf{v}_+
\exp\Bigg(2\int_0^t\Lambda_+(t^\prime)\mathrm{d}t^\prime\Bigg)
+\mathbf{v}_-\mathbf{v}_-\exp\Bigg(2\int_0^t\Lambda_-(t^\prime)
\mathrm{d}t^\prime\Bigg), \label{adiabatic}
\end{equation}
where $\mathbf{v}_\pm\mathbf{v}_\pm$ are the dyads formed from the
eigenvectors $\mathbf{v}_\pm(t)$ of $\tensor{L}(t)$. Since
$\tensor{L}$ is not normal, Eqn.~(\ref{adiabatic}) is in error by
an amount governed by $\mathbf{v}_+\cdot\mathbf{v}_-$, which is
never more than about $0.1$ and is often much smaller. Note also
that if $\sigma t_\mathrm{bend}\gg1$, then the off-diagonal
elements of $\tensor{L}$ are small compared to the diagonal
elements, causing the equations for $\hat r^{(1)}$ and $\hat
z^{(1)}$ to decouple and leading to
\begin{equation}
G(t)\approx\exp\Bigg(2\int_0^t\Lambda_+(t^\prime)\mathrm{d}t^\prime\Bigg),
\label{G-bigsig}
\end{equation}
where $\Lambda_+(t)$ is the largest eigenvalue of $\tensor{L}(t)$.
Thus, in both extremes $\sigma t_\mathrm{bend}\gg1$ and $\sigma
t_\mathrm{bend}\ll1$,
$G(t)\approx\exp(2\int\Lambda_+\mathrm{d}t^\prime)$. This result
is especially useful in the limit of small growth rate $\sigma
t_\mathrm{bend}\ll1$, since the rapid relaxation and growth of
bending modes makes it difficult to solve for $\tensor{B}$
numerically.

\begin{figure}
\includegraphics[height=4in]{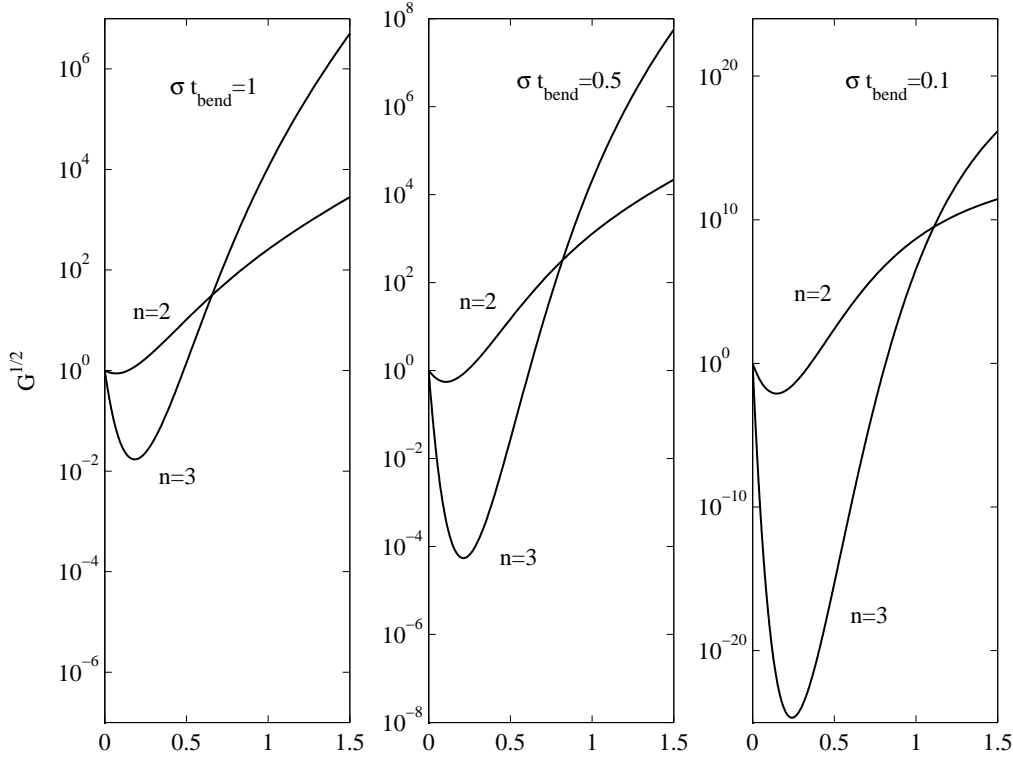}
\caption{Optimal growth factors for $\tau_0R_0=10$ and $\sigma
t_\mathrm{bend}=1.0$, $ 0.5$, $0.1$.} \label{optimal}
\end{figure}

Figure~\ref{optimal} shows the optimal growth factors for the
first two modes ($n=2$ and $n=3$) for $\sigma
t_\mathrm{bend}=1.0$, $0.5$, $0.1$ and $R_0\tau_0=10$. For both
the intermediate writhing regime and the large-$\sigma t$
asymptotic regime, the rate of increase of the growth factor
increases with mode number $n$. As $\sigma t_\mathrm{bend}$
increases, the time at which the growth factor for the $n=3$ mode
overtakes that of the $n=2$ mode decreases because the instability
of each mode occurs at earlier times. For the larger values of
$\sigma t_\mathrm{bend}$, the $n=3$ mode overtakes the $n=2$ mode
before the amplitudes have grown large enough for nonlinearities
to come into play. Thus, we expect to see double-stranded
plectonemic braids with two hairpin turns for small $\sigma
t_\mathrm{bend}$, and braids with three or more hairpin turns when
$\sigma t_\mathrm{bend}$ is large. The numerical computations of
section~\ref{numeric} confirm these expectations.

\subsection{Numerical Solution of the Nonlinear Equations}
\label{numeric} 
\begin{figure}
\includegraphics[height=3.0in]{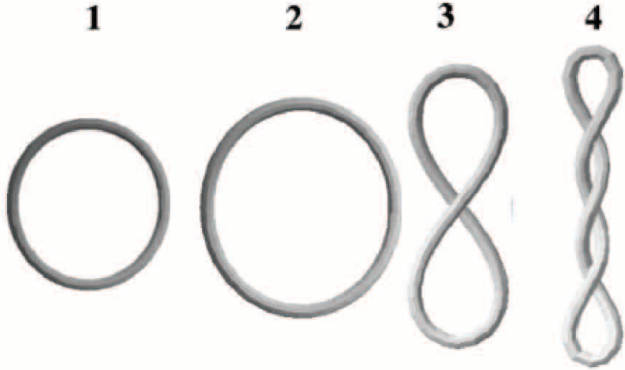}
\caption{Shape of the growing loop for $\tau_0 R_0 = 10.0$,
$\Omega_3(t=0)=8.3$,
 and $\sigma t_\mathrm{bend}= 0.1$ at (1) $t/t_\mathrm{bend}=0.03$, (2)
$t/t_\mathrm{bend}=2.41$, (3) $t/t_\mathrm{bend}=3.22$, and (4)
$t/t_\mathrm{bend}=3.37$.
} \label{energy_shapes}
\end{figure}

\begin{figure}
\includegraphics[height=2.718in]{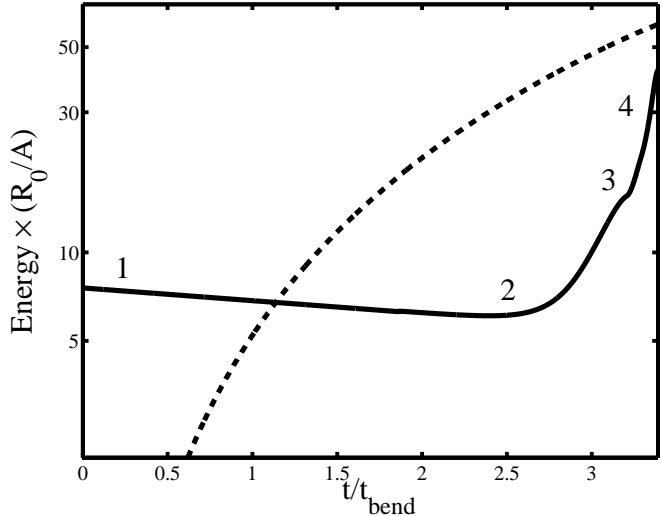}
\caption{Semi-log plot of the bend (solid
line) and twist (dashed line) energy vs. time.  Numbers correspond to
shapes in Fig.~\ref{energy_shapes} and label
important times during growth (See text for further explanation).
} \label{loopE}
\end{figure}

We solved the closed set of equations
(\ref{r-eqn},\ref{Lambdaeqn},\ref{twistdyn}) using a
pseudospectral method \cite{pseudospectral} for the backbone
dynamics (\ref{r-eqn}), direct integration of (\ref{Lambdaeqn}) at
each time step using finite differences to find $\Lambda$, and a
Crank-Nicholson routine for the twist dynamics (\ref{twistdyn}).
We used initial conditions such that the backbone of the loop is
perturbed from circular shape with $R_0=1$ by a few
small-amplitude modes ($n=2$--$5$). Depending on the simulation,
$\Omega_3$ ranged from somewhat less than $\tau_0$ to $\tau_0$.
Fig.~\ref{energy_shapes} shows a time series of the shape of the
growing loop with $\tau_0 R_0 = 10$, $\Omega_3(t=0)=\tau_0$, and
$\sigma t_\mathrm{bend} = 0.1$. For early times
(Fig.~\ref{energy_shapes}-1), the circular loop is stable and
perturbations decay. As the loop grows, $R$ increases
exponentially and $\Omega_3$ decreases.
At a critical value of $\Omega_3$ and $R$
(Fig.~\ref{energy_shapes}-2), the loop begins to buckle and wrap
about itself.  For sufficiently large $\tau_0R_0$, the loop takes
on the conformation of a {\it plectoneme}, initially forming a
figure-eight structure (Fig.~\ref{energy_shapes}-3) and then
wrapping into a braided form (Fig.~\ref{energy_shapes}-4).
Fig.~\ref{loopE} shows the twist energy, $\int
C(\Omega_3-\tau_0)^2/2 ~\rm{d}s$, and bend energy, $\int
A\kappa^2/2 ~\rm{d}s$, for the growing loop depicted in
Fig.~\ref{energy_shapes}. Note that the total energy is not fixed
in our model since growth acts to inject energy into the system.
At point 1 in Fig.~\ref{loopE}, growth along the filament axis
leads to a decrease of twist in time and thus an increase in the
twist energy. At the same time, backbone perturbations die away
and the curvature decreases exponentially, leading to a decrease
in the bend energy.  At point 2, the circular loop becomes
observably unstable. The bend energy begins increasing as
perturbations in the filament grow, and the twist energy continues
increasing (See Fig.~\ref{loopE}). At the inflection point, point
3, the filament forms a figure-eight pattern.  Note that a
figure-eight shape of a closed loop which is not growing is a
minimum of the total energy for a range of twist. At later times
(such as point 4), the filament wraps into a braided structure.
The bend energy increases as more braids are added. The twist
energy also increases; however, writhing motions act to decrease
the twisting stress imposed by growth, leading to a twist energy
that grows sub-exponentially (see Fig.~\ref{loopE}).

\begin{figure}
\includegraphics[height=3.918in]{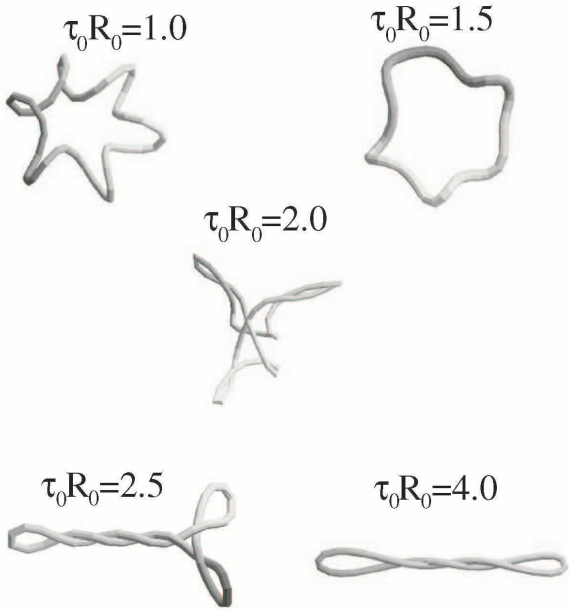}
\caption{Morphology of the growing loop at different values of
$\tau_0 R_0$.  Solenoidal morphologies occur when $\tau_0 R_0 < 2$.
Branched plectonemes are observed for $2 \le \tau_0 R_0 \le 3$ and unbranched
plectonemes are observed when $\tau_0 R_0 > 4$.  $\sigma t_\mathrm{bend}
= 0.01$ for all figures. } \label{tau0_loop}
\end{figure}

The dynamic equations were solved for a range of $\tau_0 R_0$ and
$\sigma t_\mathrm{bend}$. When $\tau_0 R_0$ was larger than about
4--10 (the actual value depended on the value of $\sigma
t_\mathrm{bend}$), we found plectonemes. Smaller values of $\tau_0
R_0$ yielded solenoids. Figure~\ref{tau0_loop} shows how the
morphology of the loop depends on the value of $\tau_0 R_0$ at
fixed $\sigma t_\mathrm{bend}$. In the region with plectonemes,
the shapes form as a result of a writhing instability; the
Euler-like buckling instability~\cite{shelleyUeda,shelley,drasdo}
plays little role. In particular, the braided shapes remain after
growth ceases. Thus, the shapes are qualitatively similar to the
minimizers of the elastic energy without growth. However, as
discussed in section~\ref{lin_stab_anal}, the rate of growth
affects the shape since the number of branches increases with
$\sigma t_\mathrm{bend}$. In the region with solenoids, the
Euler-like buckling instability comes into play since the small
value of $\tau_0R_0$ delays the onset of the writhing instability.
Solenoids are not minimizers of the elastic energy without growth;
when growth ceases the solenoids relax. Therefore, the solenoids
are the three-dimensional analogs of the two-dimensional shapes of
references~\cite{shelleyUeda,shelley,drasdo}. The plectonemic,
three-armed, and solenoidal morphologies we obtain are similar to
many of the supercoiled patterns that are observed in {\it B.
subtilis}; in the conclusion we discuss the relation between our
results and the experiments.

\begin{figure}
\includegraphics[height=3.718in]{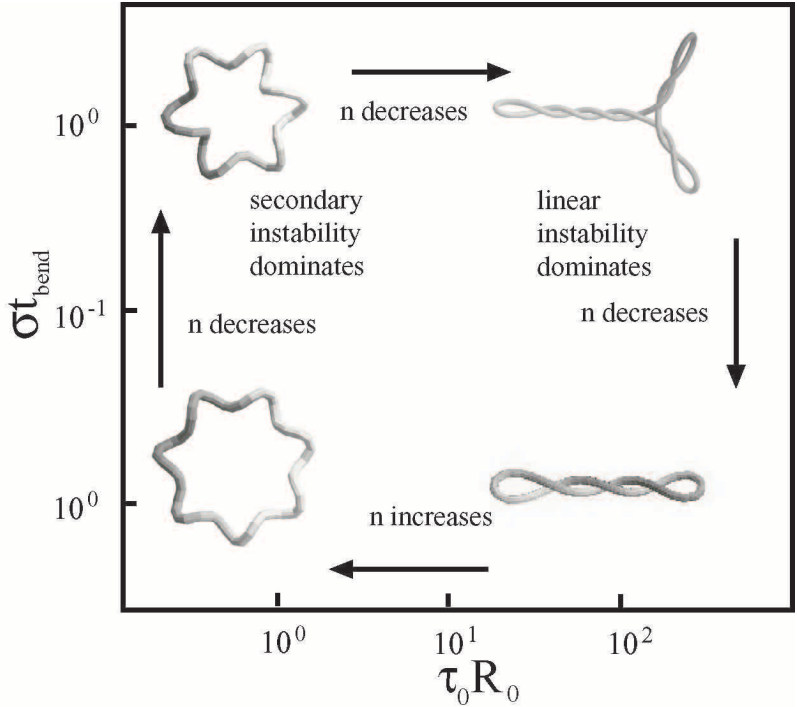}
\caption{Qualitative phase diagram depicting the dominant mode
behavior as a function of $\sigma t_\mathrm{bend}$ and $\tau_0
R_0$.  For $\tau_0 R_0$ greater than roughly 10, the linear
stability analysis predictions are valid and the $n=2$ and $n=3$
modes dominate the pattern formation.  At lower values of $\tau_0
R_0$, a secondary instability drives pattern selection and higher
order modes dominates.  Insets show the morphology of the loop in
different regions of the diagram. } \label{loop_phase}
\end{figure}

In these calculations, we defined the time to the onset of the
instability as the point where the bend energy begins increasing
(point 2 in Fig.~\ref{loopE}). This time should correspond roughly
to $t_2$ defined in section~\ref{lin_stab_anal}, since the
perturbations must be noticeably large before they affect the
bending energy. Just as in the linear stability analysis,
increasing $\tau_0 R_0$ or $\sigma$ reduces the time to onset of
the instability. For values of $\tau_0 R_0 \ge 10$, the time to
onset found numerically was in quantitative agreement with the
linear stability analysis (see Fig.~\ref{tcrit}).  At smaller
values of $\tau_0 R_0$ the onset time from the simulations was
earlier than that predicted by the linear analysis, with the
deviation getting larger for decreasing $\sigma t_\mathrm{bend}$.
The deviation is due to a secondary instability: before the
growing $n=2$ mode becomes observably large, it is overtaken by a
higher order mode which quickly dominates the shape of the loop.
The value of the mode that dominates depends on the value of
$\sigma t_\mathrm{bend}$ and $\tau_0 R_0$. As $\tau_0 R_0$ is
decreased at constant $\sigma t_\mathrm{bend}$, the mode that
dominates increases. As $\sigma t_\mathrm{bend}$ is increased at
constant $\tau_0 R_0$, the mode that dominates decreases.
Figure~\ref{loop_phase} shows the phase diagram and examples of
plectonemes and solenoids.  As a check, we verified the secondary
instability by discretizing the full nonlinear equations
(\ref{r-eqn},\ref{Lambdaeqn},\ref{twistdyn}), and integrating with
the MATLAB routine ode15s (a variable order, variable time step method
for stiff problems).

\section{Conclusions}
\label{conclude}

The three essential elements of any growth process appear in our
model for {\it B. subtilis} fibers in a simple form: the mode of
growth is exponential extension with rotation, the material
properties are described by the bending and twisting elasticity of
a slender filament, and the interaction with the external
environment is governed by resistive-force theory. Thus, our model
is ideal for illustrating the basic phenomena of the physics of
growth. We have assumed that the bacterial filaments can be
treated as perfectly elastic and that the growth rates are uniform
and independent of stress. The degree to which these simplifying
assumptions hold remains an outstanding experimental problem.
Furthermore, our model does not lead to plectonemic structures for
the case of bacterial fibers with free ends for the measured
values of the growth parameters and elastic moduli. Nevertheless,
we can draw some general conclusions from our model and use our
results to suggest new mechanisms for pattern formation in the
presence of biological growth.

First, even though the ultimate microscopic mechanism for the
supercoiled patterns of {\it B. subtilis} fibers is unknown, the
blocked rotation mechanism we study here must play a role in the
supercoiling of closed fibers. For example, since the closed loops
form supercoils at lengths which are much shorter than the length
at which the open fibers supercoil~\cite{mendelson95}, we expect
that the blocked rotation mechanism dominates over whatever
mechanism causes the open fibers to supercoil. Thus, we can
directly compare our calculations with the experiments on the
supercoiling of closed filaments. Recalling the parameters of
section~\ref{buck-writhe}, we estimate $\sigma t_\mathrm{bend} \ll
1$, so that we expect unbranched plectonemes when $\tau_0 R_0
\gtrsim 4$, branched plectonemes when $2\lesssim\tau_0 R_0
\lesssim 4$, and solenoids when $\tau_0 R_0 \lesssim 2$. Typical
observations of the filaments yield unbranched plectonemes, but
branched plectonemes and solenoids also arise, depending on the
growth medium. Using the observed buckling radius for an
unbranched plectoneme, in section~\ref{buck-writhe} we were able
to estimate $\tau_0\approx10^{4}$ cm$^{-1}$, suggesting the
presence of a structure in the cell wall with pitch
$P\approx10^{-4}$ cm. The observations of helical actin-like
polymers in the cell wall with comparable pitch~\cite{errington}
support our estimate, and further suggests a starting point for a
theory for the microscopic mechanism of the supercoiling.

The second major conclusion of our work is the dynamical nature of
the pattern selection. For large $\tau_0 R_0$, the plectonemes
remain once growth ceases, and are qualitatively similar to the
minimizers of the elastic energy, although the rate of growth
plays an important role in determining the shape. For small
$\tau_0 R_0$, the solenoids are transient structures which relax
away when growth halts. Since the curvature of the bacterial
fibers can become permanent~\cite{neil}, the solenoidal shapes may
act as a template for patterns which remain in the absence of
growth. This mechanism of pattern formation via the hardening of
transient structures formed from the interplay of flexibility and
external friction may apply to other biological systems, such as
those studied in~\cite{drasdo}.

To help justify, refine, or rule out our model, we suggest three
basic experiments. First, the shape of a growing loop as function
of time should be measured precisely enough to compare with our
theory. Although our numerical results (Fig.~\ref{energy_shapes}) are
qualitatively similar to the experimentally observed shapes
(Fig.~\ref{expt_loop}), the lack of detailed information about
the evolution in time of a single loop in~\cite{mendelson76}
prevents a stringent test of our theory. Second, the change in
elastic properties with time during growth should be
quantitatively measured. For example, to what degree and how long
must a fiber be bent to develop a permanent curvature? Finally,
the nature of the twist stress in a growing fiber with free ends
should be determined. Is there a twist moment on the cross
sections of the growing fibers with free ends, leading to a
writhing mechanism qualitatively similar to the blocked rotation
mechanism for closed loops, or is the mechanism for open fibers
completely different? Future progress toward understanding the
pattern formation of {\it B. subtilis}, both at the microscopic
level of the structure of the cell wall and the more macroscopic
level treated here, depends critically on new experiments such as
these.

We thank K. Breuer, D. Coombs, A. Goriely, M. Kim, and C.H.
Wiggins for useful discussions. We are particularly grateful to
N.H. Mendelson for discussions and the micrograph of
Fig.~\ref{expt_loop}, and A. Goriely for alerting us to
references~\cite{michell} and~\cite{basset}. CWW was partially
supported by the NSF Postdoctoral Fellowship in Microbial Biology.
REG acknowledges support from NSF DMR-9812526; TRP acknowledges
support from NSF CMS-0093658.


\end{document}